\documentstyle[12pt,epsf]{article}
\newcommand{\be}{\begin{equation}}
\newcommand{\ee}{\end{equation}}
\newcommand{\ba}{\begin{array}}
\newcommand{\ea}{\end{array}}

\newcommand{\lsim}{\buildrel < \over {_\sim}}
\newcommand{\gsim}{\buildrel > \over {_\sim}}

\def\znbb{$0\nu\beta\beta$~}
\def\bbmass{$0\nu\beta\beta$-mass~}

\def\be{\begin{equation}}
\def\ee{\end{equation}}
\def\bea{\begin{eqnarray}}
\def\eea{\end{eqnarray}}

\begin{document}


\centerline{\Large{\bf
Neutrino  Mass Spectrum and}}

\centerline{\Large{\bf
Neutrinoless Double Beta Decay 
}}

\vskip 0.5cm 

\bigskip
\centerline{H.V. Klapdor--Kleingrothaus, H. P\"as}

\centerline{
\it Max--Planck--Institut f\"ur Kernphysik,}

\centerline{\it
P.O. Box 103980, D-69029 Heidelberg, Germany}
\medskip

\centerline
{A. Yu. Smirnov}

\centerline{
{\it The Abdus Salam International Center of Theoretical Physics,}}

\centerline{\it
Strada Costiera 11, Trieste, Italy}
 
\centerline{\it
Institute for Nuclear Research, RAS, Moscow, Russia} 
\medskip



\vskip 1cm

\begin{abstract}
\noindent
The  relations between the effective  Majorana mass of
the electron neutrino, $m_{ee}$, responsible  for
neutrinoless double beta decay, and the neutrino oscillation 
parameters are considered. We show that for any specific oscillation pattern
$m_{ee}$  can take any value (from zero to the existing upper
bound) for normal mass hierarchy and it can have a minimum for inverse 
hierarchy.
This means that oscillation experiments cannot fix in general $m_{ee}$.  
Mass ranges for $m_{ee}$ can be predicted in terms of oscillation 
parameters with
additional assumptions about the level of degeneracy and the type of hierarchy 
of the neutrino mass spectrum. 
These predictions for $m_{ee}$ 
are systematically studied
in the specific 
schemes of neutrino mass and flavor which explain the solar and atmospheric
neutrino data. 
The   contributions from individual mass eigenstates 
in terms of oscillation parameters have been quantified. 
We study the dependence 
of $m_{ee}$ on the non-oscillation parameters: the overall scale of
the neutrino mass and the relative mass phases. 
We analyze  how forthcoming oscillation experiments 
will improve the predictions for $m_{ee}$. 
On the basis of these studies we evaluate the discovery potential of
future \znbb decay searches. The role 
\znbb decay searches will play in the reconstruction of the neutrino mass
spectrum is clarified. The key scales of $m_{ee}$, which will lead to the
discrimination among various schemes are: $m_{ee} \sim 0.1$ eV 
and $m_{ee} \sim 0.005$ eV. 

\end{abstract}

\section{Introduction }

The  goal of  the search for neutrinoless double beta decay 
(\znbb decay) is to
establish the violation of (total) lepton number $L$ and 
to measure the Majorana mass  of the electron neutrino, thus 
identifying  the nature of the neutrino \cite{doi85,bbrev}. 
Both issues are related: Even if the main mechanism of  
\znbb decay may be induced by e.g. lepton number violating  right-handed 
currents, R-parity violation in SUSY models, 
leptoquark-Higgs couplings (for an overview see  e.g. \cite{kk}),
the observation of \znbb decay implies always a 
non-vanishing effective  neutrino Majorana mass (\bbmass) at loop level  
\cite{schech}.

If \znbb decay is induced dominantly by the exchange  of
a  light Majorana neutrino ($m < 30$ MeV),  
the  decay rate  is  proportional to the 
Majorana mass  of the electron neutrino $m_{ee}$ squared:
\be
\Gamma \propto m_{ee}^2.
\ee
Thus, in absence of lepton mixing the observation of \znbb decay
would provide an information about the absolute scale  of  
the Majorana neutrino mass.

The situation is changed  in presence of neutrino mixing 
when the electron neutrino is not a mass eigenstate but turns out to be 
a combination of several mass eigenstates, $\nu_i$, with mass 
eigenvalues $m_i$:  
\be
\nu_e = \sum_i U_{ei} \nu_i~, ~~~~     i = 1, 2, 3,... ~.  
\ee
Here $U_{ej}$ are  the elements of the mixing matrix 
relating the flavor states  to the mass eigenstates.  
In this general case  the mass parameter (\bbmass) which enters 
the \znbb decay rate is not the physical mass of the neutrino 
but the combination $|m_{ee}|$ of physical masses: 
\be
|m_{ee}| = \left|\sum_j |U_{ej}|^2 e^{i\phi_j}m_j\right|.
\label{obs}
\ee
Apart from the absolute values of masses $m_j$ and mixing matrix elements,  
the effective Majorana mass depends also on new parameters: 
phases $\phi_j$ which originate from a possible complexity  of the mass
eigenvalues and from the mixing matrix elements. 
Thus searches for double beta decay are sensitive   
not only to  masses but also to   mixing matrix elements   
and  phases $\phi_j$.

Notice that in the presence of mixing  $m_{ee}$ is still  the $ee$-element
of the neutrino mass matrix in the flavor basis 
\footnote{In general 
the experimental value of $m_{ee}$ depends on the process 
being
considered. It coincides with the theoretical $m_{ee}$ of eq. (\ref{obs})
if all
masses $m_i \ll Q$, where $Q$ is the energy release of a given process. 
This fact may become important for comparing heavy 
neutrino contributions in \znbb decay 
and inverse neutrinoless double beta decay
at colliders, see {\it e.g.} \cite{Lon99}.}.  
In this sense it gives the scale of elements of the neutrino mass
matrix. 
However, in general,  $m_{ee}$ does not determine the scale of the 
physical  masses. If 
\znbb decay 
will be discovered and if it will be proven to proceed via the Majorana
neutrino mass mechanism, then the
$m_{ee}$ extracted from the decay rate will give 
a lower bound on some physical masses.  
As it is easy to see from Eq. (\ref{obs})),  at least   
one  physical mass, $m_j$,   should be   
\be 
m_j \geq m_{ee}
\ee
for the three-neutrino case.

Can  $m_{ee}$ be predicted? 
According to (\ref{obs}) the mass  $m_{ee}$ depends on absolute values of   
masses, mixings and  phases $\phi_j$. 
Certain  information about masses and mixing can  be obtained from
(i)  oscillation searches, (ii) direct kinematical measurements and 
(iii) cosmology.   Let us comment on these issues in order.

1).  The oscillation pattern is determined by 
mass squared differences,  
moduli of  elements of the mixing matrix, and (for three
neutrino mixing)  
only one complex phase which leads to   CP violating effects 
in neutrino oscillations:  
\be 	
\Delta m_{ij}^2 \equiv |m_i|^2 - |m_j|^2 ,~~~~ |U_{ej}|^2, ~~~~
\delta_{CP} . 
\label{oscpar}
\ee
(We indicated here only  mixing elements which enter $m_{ee}$.)     	
In what follows we will call (\ref{oscpar}) the  
{\it oscillation parameters}. 
  
Neutrino oscillations and neutrinoless double beta decay,  
however, depend on different combinations of neutrino masses and mixings.
In terms of the oscillation parameters the mass 
(\ref{obs}) can be rewritten as 
\be
|m_{ee}| = \left|\sum_j |U_{ej}|^2 
e^{i\phi_j} \sqrt{\Delta m^2_{j1} + m_1^2} \right|, 
\label{obs1}
\ee
where we assumed for definiteness $m_1$ to be the smallest 
mass. We also put $\phi_1 = 0$ and consider the other $\phi_j$ 
as {\it the relative phases}.  

According to (\ref{obs1}) the oscillation parameters do not allow one 
to determine uniquely $m_{ee}$. 
Apart from these parameters, 
the mass $m_{ee}$ depends also on the 
absolute value of the first mass (absolute scale) and on the relative
phases: 
\be 
m_1, ~~~~~ \phi_{j}, ~~~~~j = 2, 3, ...
\label{phases}
\ee
These parameters can not be determined from oscillation
experiments and we will call them {\it non-oscillation parameters}. 

The mass squared difference gives 
the absolute value of the mass only in the case of strong mass hierarchy: 
$m_j \gg m_1$, when 
$|m_j| \approx \sqrt{\Delta m_{j1}^2}$.   
However, even in this case  the lightest mass (which can  give a
significant or even dominant contribution to  $m_{ee}$) is not determined.

The relative  phases $\phi_j$ which appear in $m_{ee}$ (eq.
(\ref{obs})) differ from $\delta_{CP}$ and 
can not be determined from  oscillation experiments, 
since the oscillation pattern is determined 
by moduli $|m_i|^2$.  On the other 
hand, the phase relevant for neutrino oscillations does not enter 
$m_{ee}$ or can be absorbed in phases of masses. 


2).  Apart from neutrino oscillations, informations on neutrino masses and 
especially on the absolute scale of masses  
can be  obtained from direct kinematical searches and cosmology. 

There is still some chance that future kinematical  
studies of the tritium  beta decay  will measure the
electron neutrino mass, and thus will allow us to fix the absolute
scale of masses.  Projects are under consideration 
which will have a sensitivity of about 1 eV and less (ref. \cite{tritium}).

3). The expansion of the universe and its large  scale structure 
are sensitive to neutrinos with masses larger than about 0.5 eV. 
The status of neutrinos as the hot dark matter (HDM) component of the universe 
is rather uncertain now: it seems that the present cosmological
observations do not require 
a significant $\Omega_{\nu}$ contribution and therefore a large
${\cal O}$(1eV) neutrino mass. However in some cases massive neutrinos may 
help to
get a better fit of the data on density perturbations.  

In order to predict $m_{ee}$ one should not only determine
the oscillation parameters but make additional 
assumptions which will fix the non-oscillation parameters. 
If the oscillation parameters are known, then, 
depending on these assumptions, one can predict $m_{ee}$ completely 
or get certain  bounds on $m_{ee}$.   

What are these assumptions? 

It was pointed out in \cite{petc94} that predictions on $m_{ee}$ 
significantly depend on two points: 
\begin{itemize}

\item
The level of degeneracy of the 
neutrino mass spectrum, which is related to the absolute scale of 
neutrino masses. 

\item 
The solution of the solar neutrino problem; this solution 
determines  to a large extent the distribution of the electron neutrino
flavor in the mass eigenstates, that is, $|U_{ej}|^2$. 

\end{itemize}
 
The 
assumptions about the level of degeneracy allow one to fix the absolute scale
of the neutrino mass. 
In fact, at present even the oscillation parameters are essentially unknown,  
so that further  assumptions are needed. 
Evidences of neutrino oscillations 
(atmospheric, solar neutrino problems, LSND result) 
allow us in principle to determine the oscillation parameters up to a 
certain ambiguity related, in particular, to the existence of several 
possible solutions of the solar neutrino problem.

A number of studies of the \bbmass have been performed, using  various 
assumptions about the hierarchy/degeneracy of the spectrum  
which remove  the ambiguity in interpretations of 
existing  oscillation data. 
In fact, these assumptions allow one to construct the neutrino mass and 
mixing spectrum, and some studies have been performed for specific 
neutrino spectra. Most of the spectra considered so far 
explain the atmospheric neutrino  problem and 
the solar neutrino problem assuming one of the suggested solutions. 
Some results have also been obtained for schemes with 
4 neutrinos which also explain the LSND result. 
Let us summarize the main directions of these studies.

(1) Three-neutrino schemes with normal mass hierarchy 
which explain the solar and atmospheric neutrino data  have 
been studied in \cite{petc94,bil1,bil97,bil6,pet99,gen}. 
Various solutions of the $\nu_{\odot}$-problem were 
assumed. 
These schemes  give the  most stringent constraints on \bbmass 
in terms of oscillation parameters.

(2) The \bbmass in 
three-neutrino schemes with inverse mass hierarchy 
 has been considered in  \cite{bil2,pet99,gen}.  These schemes favor 
$m_{ee}$ to be close to the present experimental bound. 

(3) Three-neutrino schemes with {\it partial} degeneracy  of the spectrum 
and various solutions of the $\nu_{\odot}$-problem were discussed  
in \cite{petc94,gen}. In these schemes  $m_{ee}$ can also be close to the
present experimental bound.  

(4) Large attention was devoted to the three-neutrino schemes
with complete degeneracy 
\cite{petc94,adh98,min97,min97b,vis97,vis99,bar99,ell,bra99,glas,gen}  
since they can explain solar and atmospheric
neutrino data and also give a significant amount of the HDM 
in the universe. In these schemes the predictions of $m_{ee}$ depend 
mainly on  the absolute mass scale  and  on the mixing angle 
relevant for the solar neutrinos.  

%
%
%

Some intermediate situations between hierarchical and degenerate 
spectra have been discussed in \cite{vis99,gen}.

5.  The \bbmass in scenarios with  4 neutrinos which can accommodate 
also the LSND result  have been analyzed in Ref. 
\cite{bil2,bil97,bil6}. 

Some general 
bounds on the \bbmass under various assumptions have been discussed in  
\cite{fuk97n,fuk97,fuk98,vis99,vis99b,gluza}.

In a number of papers an inverse problem has been solved: using relations
between the \bbmass and oscillation parameters which appear in certain schemes 
restrictions on oscillation parameters have been found from existing bounds
on $m_{ee}$. In particular the $3\nu$-schemes with mass degeneracy 
\cite{adh98} and mass hierarchy \cite{pet99} have
been discussed. 

An 
important ingredient for the prediction of  $m_{ee}$ are the phases 
(see eq. \ref{phases}). 
Unfortunately, there is no theory or compelling assumptions which 
allow to determine these phases.

In this paper we will analyze the discovery potential of future  
\znbb decay searches in view of existing and forthcoming oscillation
experiments. We will clarify the role \znbb decay searches will play in the 
identification of the neutrino mass spectrum. 
In the previous studies, implications of $m_{ee}$
for oscillation parameters and the other way around, 
implications of oscillations searches for  $m_{ee}$  have been
discussed. In contrast,  we focus here on the impact of 
results from  both
neutrino oscillations and double beta decay on  the reconstruction of
the neutrino mass spectrum. 
We put an  emphasis on possible future experimental results from 
long-baseline experiments, CMB
explorers, supernovae measurements, precision studies of properties
of the solar  neutrino fluxes  (day-night asymmetry, 
neutrino energy spectra, etc.). 

For this we first (sect. 2)  consider the general relations between 
the effective Majorana mass of the electron neutrino and 
the oscillation parameters. We will study the dependence of 
$m_{ee}$ on the non-oscillations parameters.  
The crucial assumptions which lead to  predictions for $m_{ee}$ are
identified.

In   sects. 3 - 8
we present a systematic and updated study of predictions for 
$m_{ee}$ for possible  neutrino mass spectra. 
In contrast with most previous studies 
using oscillation data we quantify the 
contributions from {\it individual} mass eigenstates and we keep explicitly 
the dependence on unknown relative mass phases. 
The dependence of predictions on  
non-oscillation parameters -- the absolute mass  value and the phases 
$\phi_i$ is studied in detail. 
We consider  $3\nu$--schemes with mass hierarchy
 (section 3), partial degeneracy (section 4), 
total degeneracy  (section 5), transition regions (6), 
inverse hierarchy  (section 7)
and schemes with sterile neutrinos  (section 8).
We analyse how forthcoming and planned oscillation experiments  will
sharpen the predictions for $m_{ee}$. 
In sect. 9,  comparing predictions of $m_{ee}$ 
from different schemes we clarify the role  
future searches for \znbb decay  can play 
in the identification of the
neutrino mass spectrum.\\


\section{Neutrino oscillations and neutrinoless double beta decay 
\label{nobb}}

As has been pointed out in
the introduction, the prediction of $m_{ee}$
depends on oscillation ($|U_{ei}|$, $\Delta m^2$)   
 and non-oscillation ($m_1$ and $\phi_j$) parameters. 
In this section we will consider general 
relations between $m_{ee}$ and the oscillation parameters. 
We analyse the dependence of these relations on non-oscillation
parameters. 
We quantify ambiguities which exist in predictions of $m_{ee}$.  
Our results will be presented in a way which will be convenient for 
implementations of future oscillation results.

\subsection{Effective Majorana mass and oscillation parameters}

The oscillation pattern is determined by the effective 
Hamiltonian  
(in the flavor basis): 
\begin{equation}
H = \frac{1}{2E} M M^{\dagger} + V~, 
\label{hamilt}
\end{equation}
where $E$ is the neutrino energy, $M$ is the mass matrix  and 
$V$ is the (diagonal) matrix of effective potentials which describe the
interaction of neutrinos in a  medium. 

The oscillation pattern is not changed if we add to $H$ 
a term proportional to the unity matrix:   
\begin{equation}
M M^{\dagger}  \rightarrow M M^{\dagger} \pm m_0^2 I~.  
\label{add}
\end{equation}
Indeed, the additional term  
does not change the mixing, it  leads just to 
a shift of the mass eigenstates squared by the same value without
affecting  $\Delta m_{ij}^2$: 
\begin{equation}
m_i^2 \rightarrow m_i^2 \pm m_0^2~.  
\label{shift}
\end{equation} 
(we consider $m_i^2 - m_0^2 \geq 0$ for all $i$ to keep the Hermiticity 
of the Hamiltonian).    
The additional term changes, however, the \bbmass. 
Thus,  for a given oscillation pattern 
there is a freedom in $m_{ee}$,  
associated with  $m_0^2$. 

\begin{figure}[!t]
\parbox{6cm}{
\epsfxsize=60mm
\centerline{\epsfbox{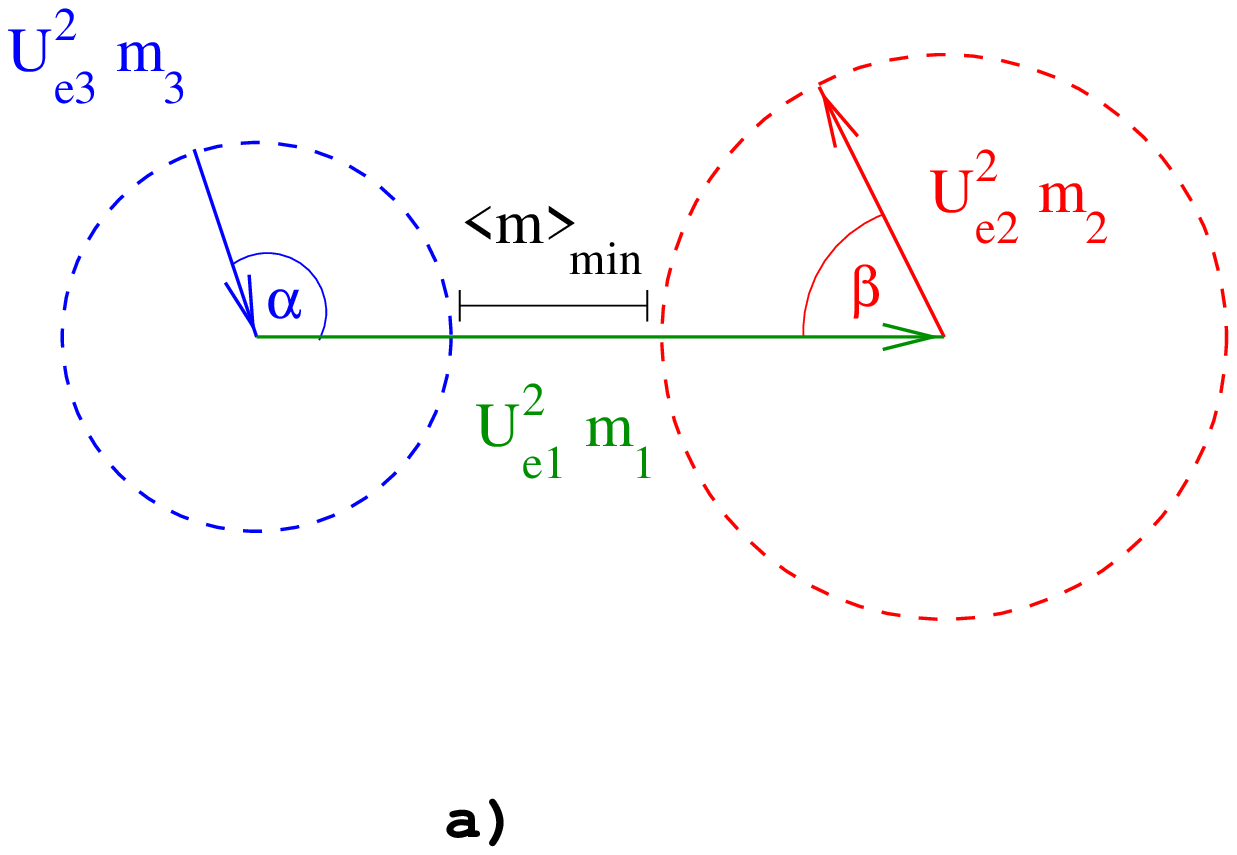}}
}
\parbox{6cm}{
\epsfxsize=60mm
\centerline{\epsfbox{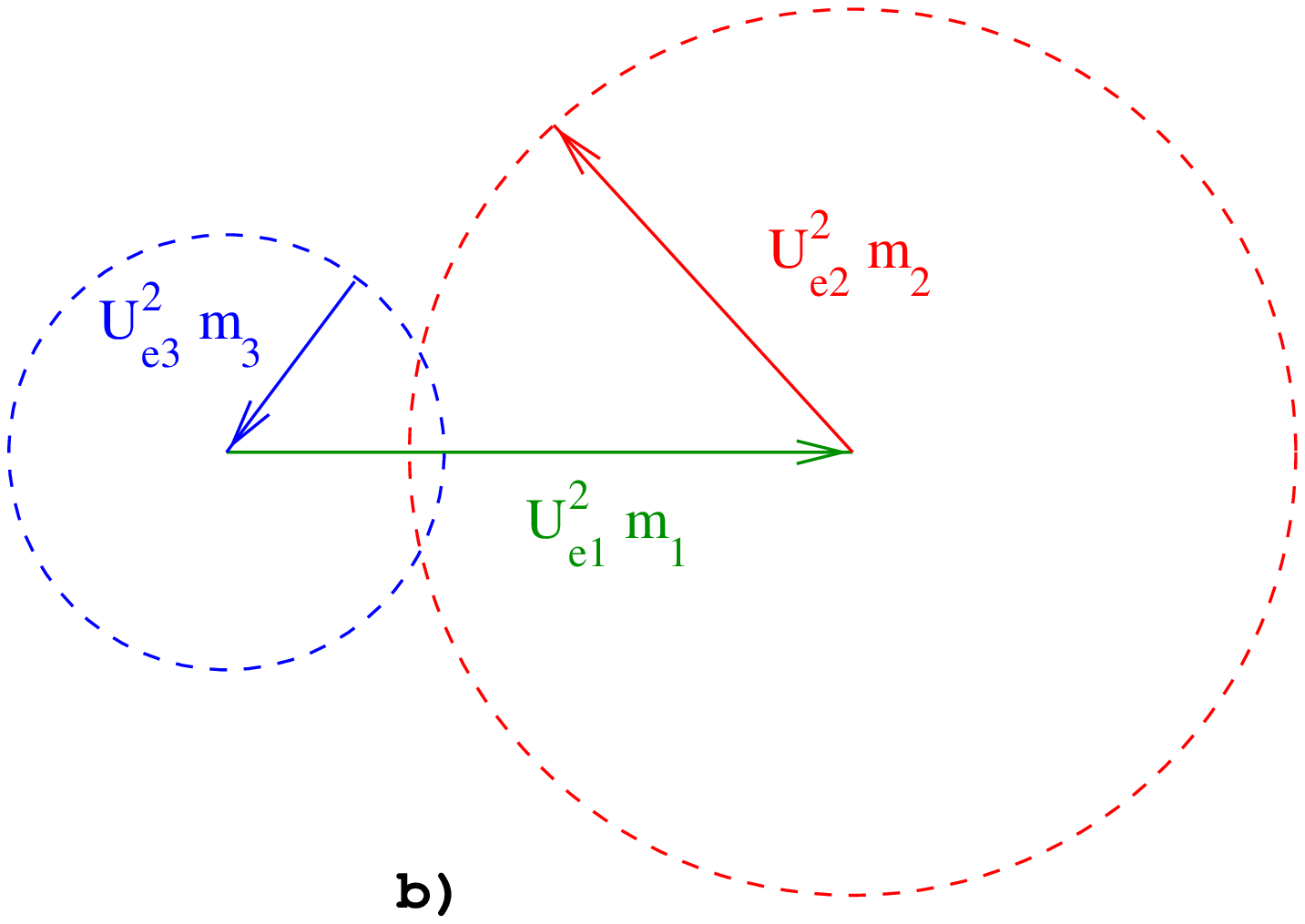}}
}
\caption{The effective Majorana mass $m_{ee}$ in the complex plane. 
Vectors show contributions to $m_{ee}$ from individual eigenstates. 
The total $m_{ee}$  appears as the sum of the three vectors. 
Allowed values of $m_{ee}$ correspond to modulies of
vectors which connect two points on the circles. Here
$\alpha = \phi_{3}-\pi$,  $\beta= \pi-\phi_{2}$.
a). $|m^{(1)}_{ee}|>|m^{(2)}_{ee}|+|m^{(3)}_{ee}|$: the 
vectors $\vec{m}^{(i)}_{ee}$ can not form a triangle 
and no complete cancellation occurs. b)   
$|m^{(1)}_{ee}|\leq|m^{(2)}_{ee}|+|m^{(3)}_{ee}|$:   
in this case complete cancellation occurs in the intersection points of
the circles, so that $m_{ee} = 0$. 
\label{graph2}}
\end{figure}

Let us  study how arbitrary $m_{ee}$ can be for a given 
oscillation pattern. 
According to (\ref{add},\ref{shift}) for the three-neutrino case we get 
\be
m_{ee} = 
|m^{(1)}_{ee}| + e^{i\phi_{2}} |m_{ee}^{(2)}|
+  e^{i\phi_{3}} |m_{ee}^{(3)}|~,
\label{mee}
\ee
where $m_{ee}^{(i)}\equiv |m_{ee}^{(i)}| \exp{(i \phi_i)}$ 
($i = 1, 2, 3$)  are  the contributions to $m_{ee}$
from individual mass eigenstates which can be written in terms 
of oscillation parameters as:  
\bea
|m^{(1)}_{ee}|&=&|U_{e1}|^2 m_1, \\
|m^{(2)}_{ee}|&=&|U_{e2}|^2 \sqrt{\Delta m^2_{21} + m_1^2},\\
|m^{(3)}_{ee}|&=&|U_{e3}|^2\sqrt{\Delta m^2_{31} + m_1^2}.
\eea
and  $\phi_{i}$  are the relative phases of the contributions from 
masses $m_i$ and $m_j$ 
(the mass $m_0^2$  has been absorbed in definition of  $m_1^2$). 
The contributions $m_{ee}^{(i)} $ 
can be shown as vectors in the complex
plane (fig. \ref{graph2}). 

Without loss of generality we assume $m_3 > m_2 > m_1 \geq 0$, so that 
$m_1$ is the  lightest state. Then {\it normal} mass hierarchy 
corresponds  to the case 
when the electron flavor prevails in the lightest state: 
$|U_{e1}|^2  > |U_{e2}|^2, |U_{e3}|^2$. We will refer to 
{\it inverse} hierarchy  
as to the case when $|U_{e1}|^2  < |U_{e2}|^2$ or/and $|U_{e3}|^2$, 
{\it i.e.}  
when the admixture of the electron neutrino flavor in the lightest state is
not the largest one.

Let us consider the dependence of $m_{ee}$ on 
{\it non-oscillation parameters}  
$m_{ee} = m_{ee}(m_1,\phi_j)$. 
It is obvious that due to the freedom in the choice of $m_1$ 
there is no upper bound for $m_{ee}$. However, in some special
cases lower bounds on $m_{ee}$ exist.

Let us start with  the two neutrino case which would correspond to 
zero (or negligibly small) $\nu_e$ admixture in  one of  
mass eigenstates,  {\it e.g.}  $|U_{e3}|$.   
We consider first the case of normal hierarchy $U_{e1}^2 > U_{e2}^2$. 
For $m_1 =  0$ the effective mass $m_{ee}$  is uniquely fixed  in
terms of oscillation parameters: 
\begin{equation}
m_{ee}^0 = 
|U_{e2}|^2 \sqrt{\Delta m^2_{21}} \equiv \sin^2 \theta 
\sqrt{\Delta m^2_{21}}~, 
\label{zerom}
\end{equation}
where 
$\sin \theta \equiv U_{e2}$.
For non-zero  $m_1$, the  maximal and minimal 
values of $m_{ee}$ correspond to 
$\phi_{2} = 0$ and $\phi_{2} = \pi$. 

\begin{figure}[!t]
\parbox{10cm}{
\epsfxsize=100mm
\centerline{\epsfbox{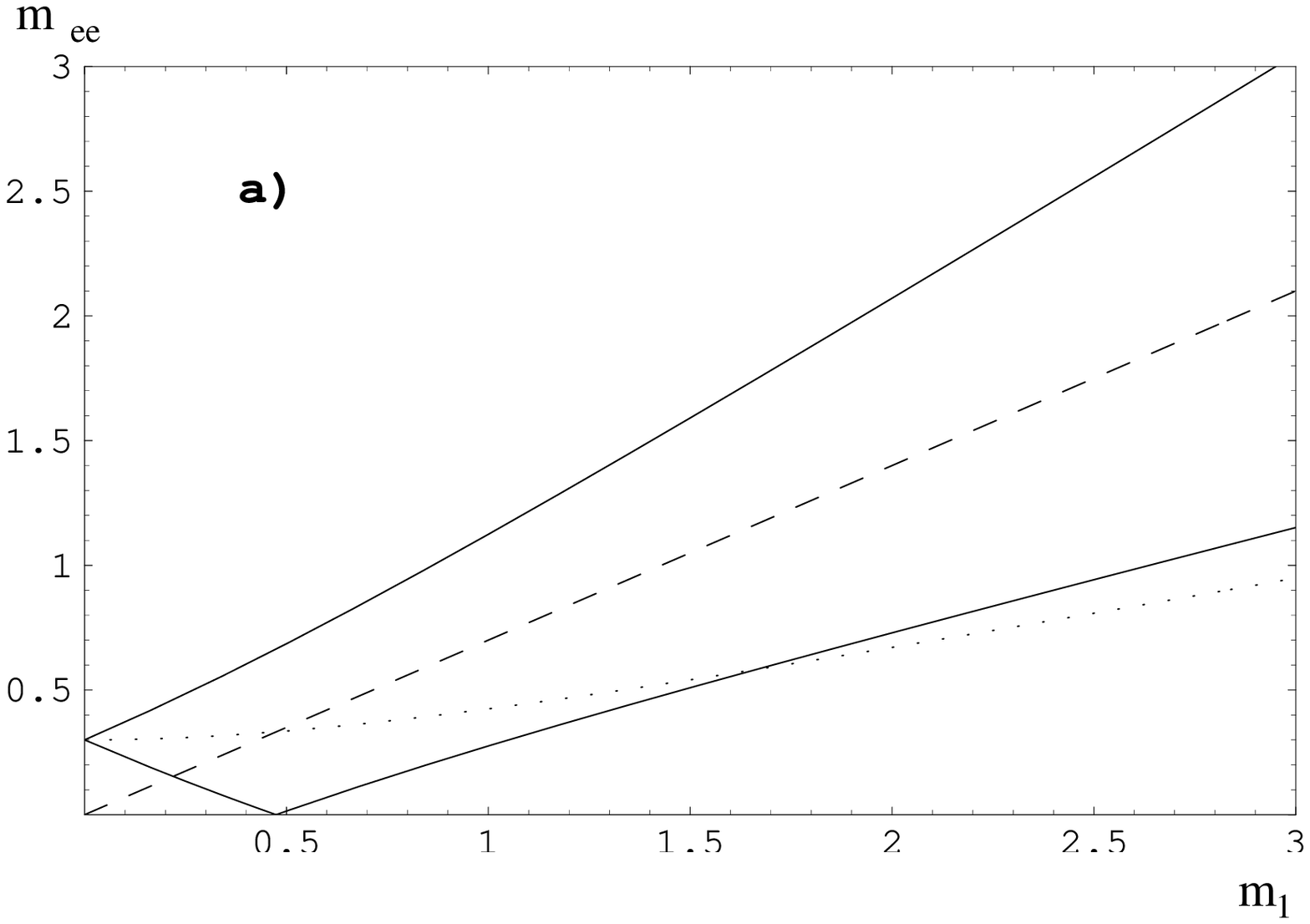}}
}
\parbox{10cm}{
\epsfxsize=100mm
\centerline{\epsfbox{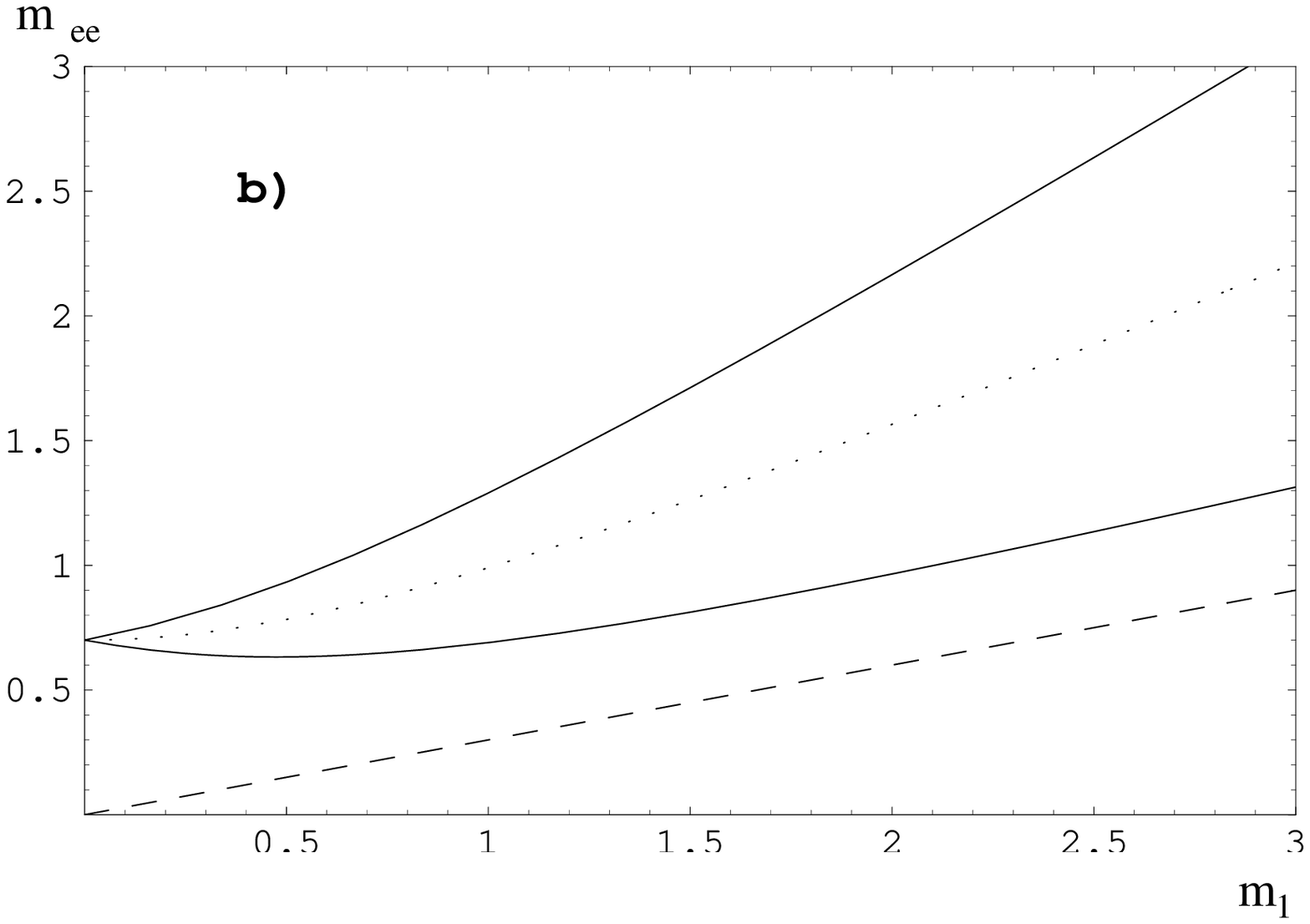}}
}
\caption{Qualitative dependence of the effective Majorana mass 
$m_{ee}$ on  $m_1$ in the two neutrino mixing case. 
a) corresponds to  the case of 
normal mass hierarchy, b) to the case of  
inverse hierarchy.
Shown are the contributions $m_{ee}^{(1)}$ (dashed) and  $m_{ee}^{(2)}$
(dotted).
The solid lines give  $m_{ee}^{max}$ and $m_{ee}^{min}$, 
which correspond to  
$\phi_{21} = 0$ and $\phi_{21} = \pi$, respectively.
While in the case of normal hierarchy a complete  cancellation is
possible, so that $m_{ee} = 0$, 
for inverse hierarchy a minimal value $|m_{ee}^{min}| > 0$ exists.
\label{g1}}
\end{figure}

The upper bound  ($\phi_{2} = 0$) 
on $m_{ee}$ increases with $m_1$ monotonously from 
$m_{ee}^0$ at $m_1 = 0$ and approaches 
the asymptotic dependence $m_{ee}  = m_1$ for large $m_1$ 
(see fig. \ref{g1} a).
The lower limit ($\phi_{2}=\pi$)  decreases monotonously 
with increase of $m_1$ starting by $m_{ee}^0$. 
It reaches zero at 
\begin{equation}
m_1 = \frac{\sin^2 \theta}{\sqrt{|\cos 2\theta|}}\sqrt{\Delta m^2_{21}}, 
\label{mass0}
\end{equation}
and approaches the asymptotic 
dependence $m_{ee}$ = $|\cos 2\theta|~ m_1$ at large $m_1$
(see fig.~\ref{g1} a).
Thus, for arbitrary values of oscillation parameters, 
no bound on  $|m_{ee}|$ exists.

(2) In the case of inverse hierarchy, $|U_{e2}| > |U_{e1}|$, 
the function  $m_{ee}(m_1)$ has a minimum which differs from zero,  
\begin{equation}
m_{ee}^{min} = \sqrt{|\cos 2\theta| \Delta m^2_{21}}
\label{minimum}
\end{equation}
at 
\begin{equation}
m_1 = \frac{\cos^2 \theta}{\sqrt{\cos 2\theta}}\sqrt{\Delta m^2_{21}}. 
\label{min3}
\end{equation}
At large $m_1$ it has the 
asymptotics $m_{ee}$ = $|\cos 2\theta| m_1$. 
(fig. \ref{g1} b).
As we will see, the existence of a minimal value of $|m_{ee}|$ can play
an important role in the discrimination of various scenarios.\\

Let us consider now the three-neutrino case. 
The mass  $m_{ee}$ is given by  the sum of
three vectors $\vec{m}_{ee}^{(1)}$, $\vec{m}_{ee}^{(2)}$ and 
$\vec{m}_{ee}^{(3)}$ in the complex
plane (see fig. \ref{graph2}), so that  
a complete  cancellation corresponds to a closed triangle. 
The sufficient condition for having a
minimal value of $|m_{ee}|$ 
which differs from zero for arbitrary non-oscillation parameters 
is  
\be
|m_{ee}^{(i)}|> \sum_{j\neq i} |m_{ee}^{(j)}|.
\label{cancelcond}
\ee
That is, one of contributions $m_{ee}^{(i)}$ should be larger than the
sum of the moduli of the two others. 

Let us  prove that this condition can not be satisfied for 
the {\it normal} hierarchy case. 
Indeed, in eq. (\ref{cancelcond}) $i$ can not be 1.  
For $m_1=0$ we have $m_{ee}^{(1)} = 0$, 
at the same time the right-handed side of eq. (\ref{cancelcond}) 
is larger than zero  as long as $U_{e1}^2 \neq 1$, 
({\it i.e.} any non-zero mixing of $\nu_e$ exists).  
The condition (\ref{cancelcond}) can not be satisfied for 
$i \neq 1$ either. In this case for a large enough $m_1$, so that 
$m_1\simeq m_2 \simeq m_3$, we get 
$m_{ee}^{(i)} < m_{ee}^{(1)}$ 
since $U_{e1}> U_{ei}$.  This proof holds also  for 
schemes with more than three
neutrinos. Thus, one can conclude that 
neither an upper nor a lower bound on $|m_{ee}|$ exists
for any oscillation pattern and normal mass  hierarchy.  
Any value $|m_{ee}| \geq 0$ can be obtained by varying the 
non-oscillation parameters $m_{1}$ and $\phi_{ij}$.\\ 

For inverse  mass hierarchy we find that 
condition eq. (\ref{cancelcond})  can be fulfilled for $i = 3$.  
Since now both $m_3> m_2, m_1$ and $|U_{e3}|>|U_{e2}|,|U_{e1}|$ one can 
get
\be
m_3 |U_{e3}|^2 > m_2  |U_{e2}|^2 + m_1 |U_{e1}|^2 
\label{canc1}
\ee
for any set of values of non-oscillation parameters provided that
the mixing of the heaviest state fulfills 
\be
|U_{e3}|^2 > 0.5. 
\label{ofive}
\ee
Indeed, for large enough $m_1$, such  that $m_1\simeq m_2 \simeq m_3$,  
the condition  (\ref{canc1}) reduces to
$|U_{e3}|^2 > |U_{e2}|^2 + |U_{e1}|^2 \equiv  1 - |U_{e3}|^2 $, 
and the latter is satisfied for (\ref{ofive}). 
For smaller values of $m_1$ the relative difference of masses   $m_3 >
m_2, m_1$ increases and the inequality of contributions 
in eq. (\ref{canc1}) becomes even stronger.
Thus, the inequality (\ref{ofive}) is the  sufficient condition for 
all values of $m_1$. 
This statement is true also for any number of neutrinos.
It is also independent of the relative size of $U_{e2}$ and $U_{e1}$.

Summarizing  we conclude that

\begin{itemize}

\item
No upper bound 
on $|m_{ee}|$ can be derived from oscillation experiments. 

\item
A lower bound exists 
only for scenarios with inverse mass hierarchy when 
the heaviest state ($\nu_3$) mixes strongly 
with the electron neutrino: $|U_{e3}|^2 > 0.5$.\\
For normal mass hierarchy certain  values of the non-oscillation
parameters $m_1$, $\phi_{j}$ exist for which $m_{ee} = 0$.

\end{itemize}

The cases of normal and inverse mass hierarchy 
(they differ by  signs of $\Delta m^2$ once the flavor of the states is fixed) 
can not be distinguished in vacuum  oscillations. 
However, it is possible to 
identify the type of the hierarchy in 
studies of neutrino oscillations in matter,  
since matter effects depend on the relative signs of the potential 
$V$ and $\Delta m^2_{ij}$. 
This will be possible  in future atmospheric neutrino experiments, 
long base-line experiments  and also studies of properties of the neutrino
bursts from supernova \cite{digh99}.

\subsection{Effective Majorana mass and the degeneracy of the spectrum}

As follows from fig. 2,  
predictions for $m_{ee}$ can be further  restricted under  
assumptions about  the absolute scale of  neutrino masses 
$m_1$. 
With increase of $m_1$ the level of degeneracy of the neutrino
spectrum increases and we can distinguish three extreme cases:

\begin{itemize}

\item
$m_1^2 \ll \Delta m^2_{21} \ll  \Delta m^2_{31}$, 
in this case the spectrum has a strong {\it  mass hierarchy}.  

\item 
$\Delta m^2_{21} \ll m_1^2  \ll  \Delta m^2_{31}$, 
this is the case of {\it  partial} degeneracy; 

\item 
Inequality $\Delta m^2_{21} \ll  \Delta m^2_{31}  \ll m_1^2$  
corresponds to {\it strong}  degeneracy.  
 
\end{itemize}

There are also two transition regions when 
$m_1^2 \sim \Delta m^2_{21}$ and 
$m_1^2  \sim \Delta m^2_{31}$. 
In what follows we will consider all  these cases in order.

\subsection{Effective Majorana mass and  present oscillation data}

Present oscillation data do not determine  precisely all 
oscillation parameters. 
The only conclusion that can be drawn with high confidence level is 
that the muon neutrino has large (maximal) mixing with 
some non-electron neutrino state. The channel 
$\nu_{\mu} \leftrightarrow \nu_{\tau}$ is the preferable one,   
and it is the only possibility,  if no sterile neutrino exists.  
Thus, in 3$\nu$ schemes the 
atmospheric neutrino data are described by  
$\nu_{\mu} \leftrightarrow \nu_{\tau}$ oscillations  
as dominant mode with  
\begin{equation}
\Delta m^2_{atm} = (2 \div 6) \cdot 10^{-3} {\rm eV}^2~,~~~
\sin^2 2\theta_{atm} = 0.84 - 1,     
\end{equation}
and the best fit point
\begin{equation}
\Delta m^2_{atm} = 3.5 \cdot 10^{-3} {\rm eV}^2~,~~~
\sin^2 2\theta_{atm} = 1.0,     
\end{equation}
\cite{Kaj99}, see also \cite{gonz2000}. 
A small contribution of the
$\nu_{\mu}\leftrightarrow \nu_e$ mode is  possible and probably required
in 
view of an excess in the $e$ -like events in the Super-K experiment.

As it was realized some time ago \cite{petc94},  predictions for 
$m_{ee}$ depend crucially on the solution of the solar neutrino problem. 
The solution of the solar 
neutrino problem determines the distribution of the $\nu_e$-flavor 
in the mass eigenstates, and this affects
considerably expectations for the \bbmass. 
Up to now the unique solution is not yet identified and there are 
several possibilities \cite{Smir99}, see also \cite{gonzmsw}:  


1. Small mixing angle MSW solution with 
\begin{equation}
\Delta m^2_{\odot} = (0.4 \div 1) \cdot 10^{-5} {\rm eV}^2~,~~~
\sin^2 2\theta_{\odot} = (0.2 \div 1.2) \cdot 10^{-2}
\label{small}
\end{equation}

2. Large mixing angle MSW solution with  
\begin{equation}
\Delta m^2_{\odot} = (0.1 \div 1.5)\cdot 10^{- 4} {\rm eV}^2~,~~~
\sin^2 2\theta_{\odot} = (0.53 \div 1)
\label{large}
\end{equation}

3. Low mass MSW (LOW) solution with  
\begin{equation}
\Delta m^2_{\odot} = (0.3 \div 2.5) \cdot 10^{- 7} {\rm eV}^2~,~~~
\sin^2 2\theta_{\odot} = (0.8 \div 1)
\label{low}
\end{equation}

4. Several regions of vacuum oscillation (VO) solutions exist with 
\begin{equation}
\Delta m^2_{\odot} < 10^{- 9} {\rm eV}^2~,~~~
\sin^2 2\theta_{\odot} > 0.7~. 
\label{VO}   
\end{equation}

There is a good chance that  before the new generation 
of double beta decay experiments starts operation
 studies of the solar neutrino fluxes by existing and forthcoming 
experiments will allow us to identify  the 
solution of the solar neutrino problem. 
The key measurements include the  
day-night effect, the zenith angle dependence of the signal during the night, 
seasonal variations,  energy spectrum distortions and the
neutral current event rate.

The LSND result \cite{lsnd} which implies 
\be
\Delta m^2_{LSND} = (0.2 \div 2)~ {\rm eV}^2,~~~
\sin^2 2\theta_{LSND} = (0.2 \div 4) \cdot 10^{-2}  
\ee
is considered as the most ambiguous hint for neutrino
oscillations. The KARMEN \cite{karmen} 
experiment does not confirm the LSND result but 
it also does not fully exclude this 
result (see \cite{lsnd}). 
The oscillation interpretation of the LSND result  will be
checked by the MINIBOONE \cite{miniboone} experiment.   
A simultaneous explanation of the LSND result  
and of the  solutions of the solar and
atmospheric neutrino problems in terms of neutrio mass and mixing
requires the  introduction of a forth neutrino. 
We will discuss the 4$\nu$ schemes  in section 8.\\

Summarizing, there is a triple uncertainty affecting predictions of  
$m_{ee}$: 

1. An uncertainty in oscillation parameters. 
The oscillation pattern does not determine uniquely the 
\bbmass. Moreover, not all relevant oscillation parameters are known, 
so that additional  assumptions are needed.

2. An uncertainty in the absolute scale $m_1$.  
Some  information on $m_1$ can be obtained from cosmology  
and may be from direct kinematical measurements.  

3. An uncertainty in the relative phases. 
Clearly, the dependence on the phases is small in the case 
if  one of the eigenstates gives a dominating contribution to 
$m_{ee}$.

In what follows we will consider 
predictions for the \bbmass in  
schemes of neutrino masses and mixings
which explain the solar and the atmospheric neutrino data. 
The schemes differ by the solution of the solar neutrino problem,   
the type of the hierarchy and the level of degeneracy. 
Relative phases are considered as free 
parameters.

\section{Schemes with normal mass hierarchy}

In the case of strong mass hierarchy, 
\begin{equation}
m_1^2 \ll \Delta m^2_{21} \ll  \Delta m^2_{31}~, 
\end{equation}
the absolute  mass values of the two heavy neutrinos are completely 
determined by the mass squared differences:  
\begin{equation}
m_3^2 = \Delta m^2_{31} =  \Delta m^2_{atm},~~    
m_2^2 = \Delta m^2_{21} = \Delta m^2_{\odot}.    
\end{equation}
The only freedom left is the choice of the value of $m_1$. 
In this case the \bbmass~ is to a large extent  
determined by the oscillation parameters. 

Since the heaviest neutrino has a mass $m_3 \leq 0.1$ eV,  
the neutrino  contribution to the Hot Dark Matter
component of the universe is
small:  $\Omega_{\nu} < 0.01$. This neutrino contribution cannot be seen
with present and future experimental sensitivity
in the CMB radiation, unless a large  lepton asymmetry 
exists \cite{pastor}. Oberservational evidence of a significant 
amount of the HDM component 
$\Omega_{\nu} \gg 0.01$ would testify against  this scenario.  

\subsection{Single maximal (large) mixing}
\label{smm}

\begin{figure}[!t] \hbox to
\hsize{\hfil\epsfxsize=7cm\epsfbox{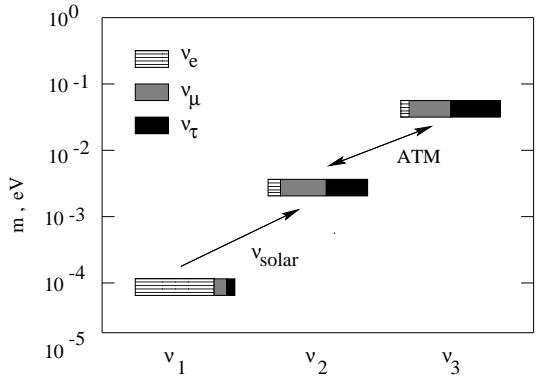}\hfil} 
\caption{~~Pattern of neutrino masses 
and mixing  in the scenario with mass hierarchy  and SMA solution  
of the  solar neutrino problem. The
boxes correspond to mass eigenstates, the sizes of different
regions in the boxes show admixtures of different flavors. Weakly hatched
regions correspond to the electron flavor, strongly hatched regions depict
the muon flavor, black regions present the tau flavor.  } 
\label{solat}
\end{figure}

In this scheme
$\nu_{\mu}$ and $\nu_{\tau}$ are mixed strongly in $\nu_{2}$ and
$\nu_{3}$ (see fig. \ref{solat}). The electron flavor is weakly mixed:  
it is mainly
in $\nu_{1}$ with small admixtures in the heavy states. 
The solar neutrino data are explained by  
$\nu_e \rightarrow \nu_{\mu}, \nu_{\tau}$ resonance conversion inside
the Sun. 
(Notice that $\nu_e$ converts
to $\nu_{\mu}$ and $\nu_{\tau}$ in comparable portions.)   
A small admixture of $\nu_e$ in $\nu_{3}$ can lead to resonantly
enhanced oscillations of $\nu_e$ to $\nu_{\tau}$
in the matter of the Earth.

\begin{figure}[!t]
\epsfxsize=15cm
\centerline{\epsfbox{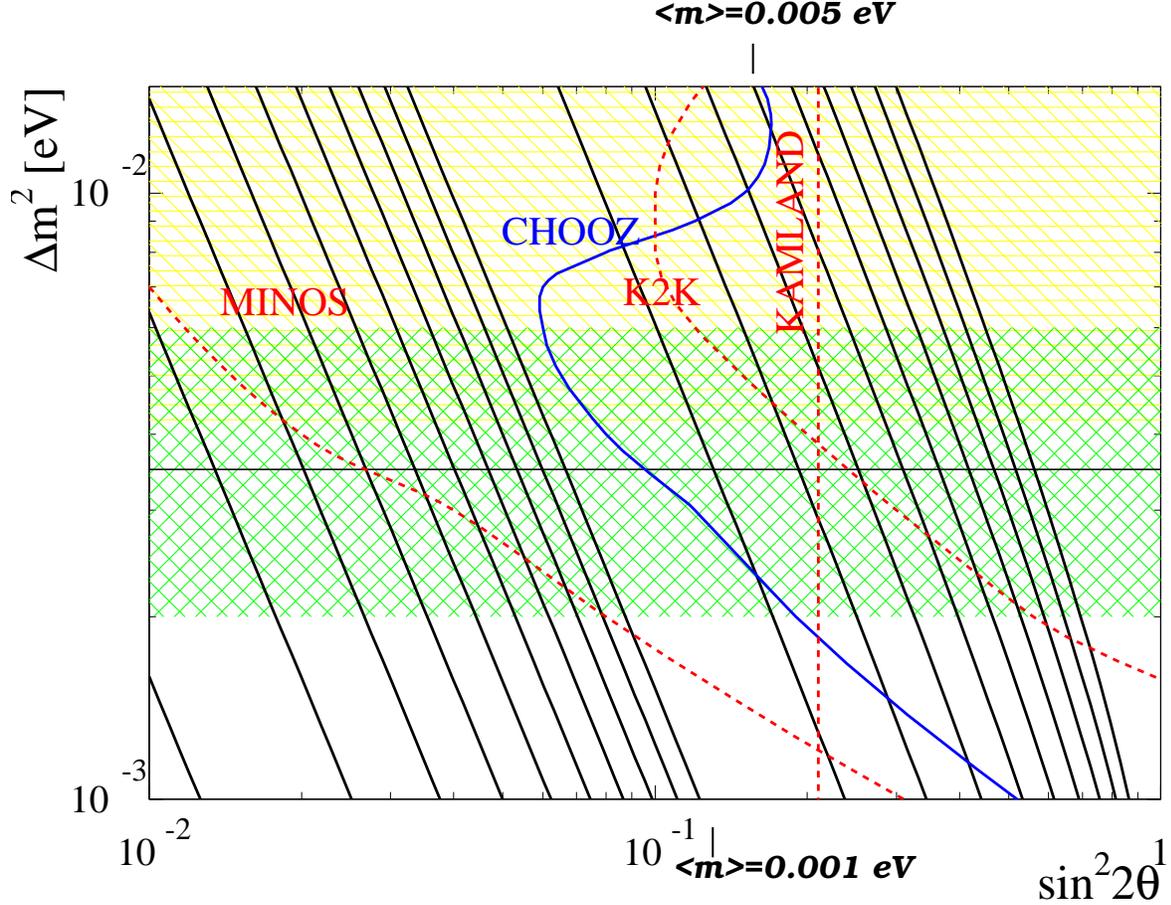}} 
\caption{
Iso-mass ($ \langle m \rangle |m_{ee}^{(3)}|$) lines (solid lines)
in the 
scheme with hierarchical mass spectrum. 
From the upper right downward $|m_{ee}^{(3)}|$ decreases from 0.01 eV  
to 0.0001 eV. Also shown are  the regions 
favored by the Super--Kamiokande atmospheric neutrino data  
with current bestfit (solid horizontal line)
and Kamiokande (lower and upper shaded areas, 
respectively,
according to \protect{\cite{Kaj99}})
and the borders of the regions excluded by  CHOOZ (solid line)
\protect{\cite{Dec99}}  
as well as the expected final sensitivity 
of KAMLAND and K2K (dashed)
\protect{\cite{Zub98}} as well as of MINOS
\protect{\cite{minos}}. 
}
\label{smix1}
\end{figure}

Let us consider the contributions to $m_{ee}$ from 
individual mass eigenstates.  
The contribution from  the third state,  $m_{ee}^{(3)}$, can be written 
in terms of oscillation  parameters as 
\be
m_{ee}^{(3)} \simeq \frac{1}{4}\sqrt{\Delta m_{atm}^2} 
\sin^2 2 \theta_{ee},
\label{third}
\ee
where the mixing $\sin^2 2 \theta_{ee} \approx 4 |U_{e3}|^2$ 
determines the oscillations of $\nu_e$ driven by the atmospheric 
$\Delta m_{atm}^2$. The parameter $\sin^2 2 \theta_{ee}$ 
immediately gives the depth of oscillation of the $\nu_e$ -survival
probability and it is severely constrained by the CHOOZ experiment. 
In fig. \ref{smix1} the iso-mass lines of equal $m_{ee}^{(3)}$ in the
oscillation 
parameter space are shown, together with  various
bounds from  existing 
and future reactor and accelerator oscillation experiments.  
The shaded region shows the favored range of 
$\Delta m_{13}^2$ from the atmospheric neutrino data.
As follows from  fig. \ref{smix1} in the Super-K
favored region the CHOOZ bound  leads to  
\be
m_{ee}^{(3)} < 2 \cdot 10^{-3} {\rm eV}. 
\label{thirdmax}
\ee 
For the best fit value of the atmospheric neutrinos
the bound is slightly  stronger: 
$m_{ee}^{(3)} < 1.5 \cdot 10^{-3}$ eV.

The mixing $\sin^2 2 \theta_{ee}$ and therefore $m_{ee}$ can  
be further restricted by  searches of $\nu_{\mu} \leftrightarrow \nu_{e}$
oscillations in the long baseline (LBL) 
experiments (K2K, MINOS, CERN-Gran-Sasso). The
effective mixing 
parameter measured in these experiments  equals 
$\sin^2 2 \theta_{e \mu} = 4 |U_{e3}|^2 |U_{\mu 3}|^2$, so that 
\be
\sin^2 2 \theta_{ee} = \frac{\sin^2 2 \theta_{e \mu}}{|U_{\mu 3}|^2}, 
\label{rel}
\ee
where the matrix element $|U_{\mu 3}|^2$ is determined by the dominant mode 
of the atmospheric neutrino oscillations. Using Eq. (\ref{rel}), 
the value $|U_{\mu 3}|^2 = 1/2$ and the expected sensitivity 
to  $\sin^2 2 \theta_{e \mu}(\Delta m^2)$ 
of K2K and  MINOS experiments, we have constructed 
corresponding bounds in  fig.~\ref{smix1}. 
According to  fig.~\ref{smix1},   
these  experiments will be able to improve the  
bound on $m_{ee}^{(3)}$ by a  factor of 2 - 5 
depending on $\Delta m^2$   and reach 
$2 \times 10^{-4}$ eV for a value of $\Delta m_{atm}^2$ at  the
present upper bound. For smaller values of $|U_{\mu 3}|^2$ the bound 
on $m_{ee}$ will be weaker. Taking  the smallest  value 
$|U_{\mu 3}|^2 = 0.3$ 
allowed by the atmospheric neutrino data, we get that the bound on $m_{ee}$ 
will be 1.7  times weaker. In any case, future LBL experiments will 
be able to probe the whole region of sensitivity of 
even the second stage of the GENIUS experiment.

A
much stronger bound on 
$m_{ee}^{(3)}$ can be obtained
from studies of neutrino bursts from Supernovae \cite{digh99}. A 
mixing parameter  as small as  
$\sin^2 2 \theta_{ee} = 10^{-4}$ can give an observable effect
in the energy spectra of supernova neutrinos. This corresponds to 
$m_{ee}^{(3)} \sim 2 \cdot 10^{-6}$ eV.

The contribution from the second mass eigenstate is completely determined by 
the parameters being responsible for the 
solution of the  solar neutrino problem:
\be
m_{ee}^{(2)} \simeq \frac{1}{4}\sqrt{\Delta m_{\odot}^2} \sin^2 2 
\theta_{\odot}.
\ee
Taking the 99 \% C.L. region of solution (\ref{small}) we obtain
\be
m_{ee}^{(2)}=(5 \cdot 10^{-7} \div 10^{-5})~{\rm eV},
\ee
and in the best fit point 
\be
m_{ee}^{(2)}=4 \cdot 10^{-6} {\rm eV}. 
\ee

The contribution from  the lightest state 
is 
\be
m_{ee}^{(1)} = m_1 \cos^2 \theta_{\odot} \simeq m_1 \ll m_2 
< 2 \cdot 10^{-3}  {\rm eV},
\ee
which can  be even larger than $m_{ee}^{(2)}$: 
if the  hierarchy between the masses of the first and the second state
is not too strong, $m_1/m_2 > 10^{-2}$   (for comparison 
$m_e/m_{\mu}=5 \cdot 10^{-3}$), we get  $m_{ee}^{(1)} > 2 \cdot 10^{-5}$
eV,  with a typical  interval 
$m_{ee}^{(1)} \sim  (0.2 - 2) \cdot 10^{-4}$ eV. \\

\begin{figure}[!t]

\epsfxsize=60mm
\centerline{\epsfbox{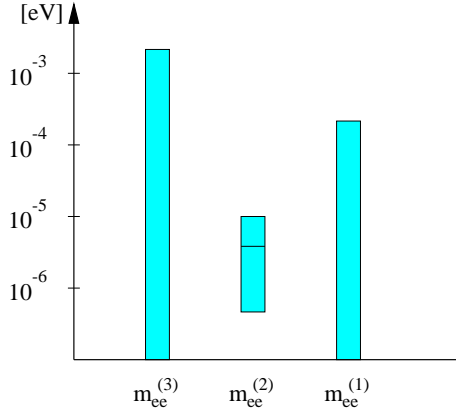}}
\caption{
Contributions to $m_{ee}$ from the individual  mass eigenstates 
for the single maximal mixing scheme with mass hierarchy. 
The bars correspond to allowed regions. 
\label{cstates1}}
\end{figure}

Summing up the contributions (see fig. \ref{cstates1}) one finds  
a maximal value for the \bbmass 
\be
m_{ee}^{max} = (2 - 3) \cdot 10^{-3} {\rm eV} 
\ee
which is dominated by the third mass eigenstate. 
No lower bound on $m_{ee}$ can be 
obtained from the present data.   Indeed,  $U_{e3}$ and therefore 
$m^{(3)}_{ee}$ can be zero.  
The same statement is true for $m_{ee}^{(1)}$, since  no lower bound
for $m_1$ exists.
The only contribution bounded from below is 
$m_{ee}^{(2)} > 10^{-6} {\rm eV}$. However, cancellations with the two  other
states can yield a zero value for the total $m_{ee}$  
(see fig. \ref{cstates1}).\\

The following conclusions on future double beta experiments and neutrino 
oscillations can be drawn

1). If future experiments will detect neutrinoless beta decay 
with a rate corresponding to $m_{ee} > 2 \cdot 10^{-3}$ eV, 
the scenario under consideration will  be excluded,  unless 
contributions to \znbb decay from  alternative mechanisms  
exist. 

2). As we have pointed out, future
long-baseline  oscillation 
experiments on $\nu_{\mu}\rightarrow \nu_e$ oscillations (MINOS)
may further improve  
the bound on $U^2_{e3}$ and therefore on $m_{ee}^{max}$ 
by a factor of $\sim 2 - 5$. 
A much stronger bound may be 
obtained from supernovae studies \cite{digh99}.
As  follows  from fig. \ref{smix1} and from the fact the SMA solution
is realized, KAMLAND should give a zero-result 
in this scheme.

3). An important conclusion can be drawn if 
future LBL and atmospheric neutrino  experiments  will observe  
 $\nu_e$-oscillations  near the present upper bound.  
In particular, an up-down asymmetry of  the $e$-like events at 
Super-Kamiokande 
is one of the manifestations of these oscillations \cite{akh99}.
In this case the $\nu_3$ contribution to $m_{ee}$ dominates, 
no significant cancellation is expected and 
the dependence on the relative phases is weak. One predicts then the result 
\be
m_{ee}  \approx m_{ee}^{(3)} \sim U_{e3}^2 \sqrt{\Delta m^2_{atm}}.
\label{pred}
\ee 
The   observation of \znbb decay with $m_{ee}^{exp} \sim m_{ee}^{(3)} $ 
would provide a strong evidence of the scheme, provided that the SMA solution 
will be established. On the other hand
it will be difficult to exclude  the scheme if 
\znbb decay will not be observed  at the level which 
corresponds to  $m_{ee}$  (\ref{pred}).  
In this case the scheme will be disfavored. However one should 
take into account also possible
cancellations of $m_{ee}^{(3)}$  and $m_{ee}^{(1)}$, 
if the mass hierarchy is weak.  


\subsection{Bi-large mixing}

The previous scheme can be modified in such a way that the  
solar neutrino data  are explained by the large angle MSW conversion. 
Now the $\nu_e$ flavor is strongly mixed in $\nu_1$ and $\nu_2$.

The contribution from the third state is the same as in 
the previous scheme  (see eq. (\ref{third})) with the upper bound   
$m_{ee}^{(3)} < 2 \cdot 10^{-3}$ eV (\ref{thirdmax}).

\begin{figure}[!t]
\hbox to \hsize{\hfil\epsfxsize=7cm\epsfbox{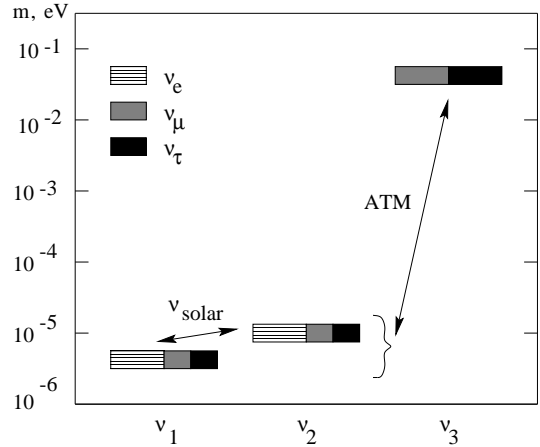}\hfil}
\caption{~~Neutrino masses and mixing pattern  of the  bi-maximal 
mixing  scheme with mass hierarchy.
}
\label{fbimax}
\end{figure}

\begin{figure}[!t]
\epsfxsize=9cm
\centerline{\epsfbox{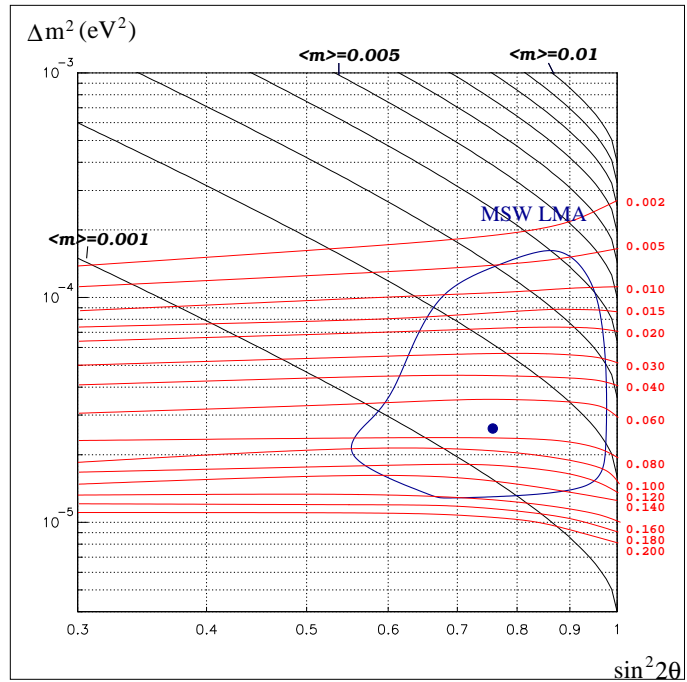}} 
\caption{
The iso-mass $\langle m \rangle = m_{ee}^{(2)}$ lines, determining the 
contribution of the second 
state in the $\Delta m_{12}^2 - \sin^2 2 \theta_{12}$ plane for the
hierarchical scheme with the LMA MSW solution.
From the upper right downward: 
$|m_{ee}^{(2)}|$ decreases from
0.01 to 0.001 eV.
Also shown is the MSW LMA 99 \% C.L. allowed region from the combined 
analysis of the Homestake, Gallex, Sage and Super-Kamiokande rates and the
Super-Kamiokande and the day-night asymmetry at Super-Kamiokande.  
The point indictes the best fit value parameters.
The horizontal lines correspond to contours of constant 
day-night assymmetry
\protect{\cite{bahneu}}. 
KAMLAND should observe a 
disappearance signal in this model.
}
\label{smix2}
\end{figure}

The contribution from the second level, 
\be
m_{ee}^{(2)}=\frac{1}{2}\left(1-\sqrt{1-\sin^2 2 \theta_{\odot}}\right)
\sqrt{\Delta m^2_{\odot}}~, 
\ee
can be 
significant: both the mixing parameter and the mass are now larger.  
According to fig. \ref{smix2}, 
in the region of the LMA solution of the solar neutrino problem, 
the contribution can vary in the interval 
\be 
m_{ee}^{(2)}=(0.5 - 4) \cdot 10^{-3} {\rm eV}~. 
\ee 
In the best fit point we get 
$
m_{ee}^{(2)}\simeq 1.4 \cdot 10^{-3} {\rm eV}.
$
Notice that a lower bound on $m_{ee}^{(2)}$ exists in this scheme,
provided that $sin^2 2 \theta<1$. Notice that a day-night asymmetry of about 
6 \%  indicated by the Super-Kamiokande experiment would correspond to
$m_{ee}^{(3)}=(1-3) \cdot 10^{-3}$ eV.

\begin{figure}[!t]
\epsfxsize=60mm
\centerline{\epsfbox{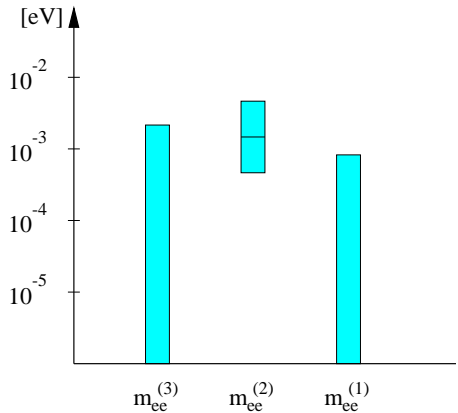}}
\caption{
Contributions from different mass eigenstates to 
$m_{ee}$ for the bi-large mixing scheme with mass hierarchy.
\label{cstates2}}
\end{figure}

The contribution $m_{ee}^{(1)}$:  
\be
m_{ee}^{(1)}\simeq \cos^2 \theta_{\odot} m_1~, 
\ee
where $\cos^2 \theta \simeq 0.5-0.84$,  
is smaller than in the previous scheme of sect. \ref{smm},  since now 
$\nu_e$ is not purely $\nu_1$ and $m_1$ can be as large as
$1 \cdot 10^{-3}$ eV for $m_1/m_2 < 0.1$. 
Due to the mass hierarchy $m_{ee}^{(1)}$ is much  smaller than 
$m_{ee}^{(2)}$ (see fig. \ref{cstates2}).

Summing up the contributions, we get a maximal value of 
$m_{ee}^{max}\simeq 7 \cdot 10^{-3}$ eV.
The typical expected  value for $m_{ee}$ is in the range of
several $10^{-3}$ eV.
However, no lower bound on $m_{ee}$ can be obtained  on the basis of the
present
data, although  values of $m_{ee}$ being
smaller than $10^{-3}$ eV require some 
cancellation of the contributions $m_{ee}^{(2)}$ and $m_{ee}^{(3)}$.\\  

Let us consider possible implications of 
future  results  from oscillations and  \znbb decay searches: 

\begin{itemize}

\item
The observation of $m_{ee}>(few)\cdot 10^{-2}$ eV  
will exclude the scheme.

\item 
The non-observation of \znbb decay  
will not exclude the scheme due to possible cancellations.

\end{itemize}

The situation can, however, change in the future, if oscillation experiments 
restrict strongly one of the contributions $m_{ee}^{(2)}$ or $m_{ee}^{(3)}$.
Let us discuss possible  developments in this direction:

\begin{itemize}

\item
Within several years solar neutrino experiments will check
the LMA-solution. In particular, further measurements of the day-night 
asymmetry and zenith angle distribution at Super-K and SNO could give a
decisive identification of
the solution of the solar neutrino problem
(see fig. \ref{smix2}). 
Notice that precise measurements of the day/night asymmetry can sharpen the
predictions of $m_{ee}^{(2)}$. 
Moreover, the LBL reactor experiment KAMLAND should observe an oscillation 
effect thus providing
an independent check of the LMA MSW solution.

\item 
If MINOS or atmospheric neutrino studies will  
fix $m_{ee}^{(3)}$ near the present 
upper bound, one can study the interference effects of 
$m_{ee}^{(2)}$ and $m_{ee}^{(3)}$ in \znbb decay 
determined by the relative phases $\phi_2$ and $\phi_3$. 
 
\end{itemize}

\subsection{Scheme with Vacuum oscillation solution}

\begin{figure}[!t]
\epsfxsize=60mm
\centerline{\epsfbox{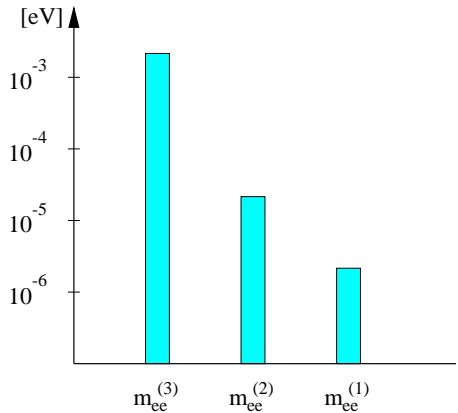}}
\caption{Contributions from different mass eigenstates to 
$m_{ee}$ for the bi-maximal mixing scheme with mass hierarchy. 
\label{cstates3}}
\end{figure}

The solar electron neutrinos $\nu_e$ oscillate in vacuum 
into comparable mixtures of  $\nu_{\mu}$ and $\nu_{\tau}$ (fig. \ref{fbimax}).
The fit to the data indicates  several disconnected regions in the
$\Delta m^2-\sin^2 2 \theta$ - plot. We  consider the 
large $\Delta m^2$ region, 
$
\Delta m^2 = (4 - 9)  \cdot 10 ^{-10} {\rm eV}^2 
$, 
and 
$ 
\sin^2 2 \theta > 0.8, 
$
where oscillations  allow one to explain an  excess of the e-like
events in the recoil electron spectrum indicated by Super-Kamiokande. 
(Obviously a small $\Delta m^2$ will give even smaller 
contributions to the effective Majorana mass). 
In this case 
\be
m_{ee}^{(2)} = \sqrt{\Delta m^2} \sin^2 \theta_{\odot} 
< 2 \cdot 10^{-5} {\rm eV}.
\ee 
Due to the mass  hierarchy and large mixing,  
the lightest mass eigenstate gives an even smaller contribution:  
$m_{ee}^{(1)}\ll m_{ee}^{(2)}$.
The contribution from the third state is the same as in  
Eq. (\ref{third}) and in fig. \ref{smix1}. For the sum we get   
\be
m_{ee} \simeq m_{ee}^{(3)} < 2 \cdot 10^{-3} {\rm eV}~, 
\ee
and clearly,  $m_{ee}^{(3)}$ can be the dominant contribution 
(fig.~\ref{cstates3}).  

The following conclusions may be drawn:
 
1). The observation of $m_{ee}>10^{-2}$  eV will exclude the scheme. 
On the other hand 
there is no minimal value for $m_{ee}$ according to the present data, so that
negative results of searches for \znbb decay will have no serious 
implications for this scheme.

2) A positive signal for atmospheric $\nu_e$ oscillations or in the MINOS 
experiment will 
allow us to predict uniquely the value of $m_{ee}$. Then searches for
$m_{ee}$ will give a crucial check of the scheme. The absolute
scale of the  neutrino mass will be fixed.\\

Similar results can be  obtained for the LOW  MSW solution. Here
the mass squared difference 
$
\Delta m_{21}^2 = 3 \cdot 10^{-7} {\rm eV}^2
$
implies 
\be
m_{ee}^{(2)}<  3\ \cdot 10^{-4} {\rm eV}.
\ee
Again $m_{ee}^{(1)}\ll m_{ee}^{(2)}$, and the main contribution may arise
from the third state.\\

Thus,  models with normal mass hierarchy  lead to rather small 
values of $m_{ee}$,  certainly  below $10^{- 2} {\rm eV}$. 
Moreover, the largest value can be obtained in the scheme with the LMA 
MSW solution of the solar neutrino problem. 
The lower bound is of the order  $\sim 10^{- 3} {\rm eV}$, 
unless cancellation (which looks rather  unnatural) occurs. 
Clearly, only the second stage of the GENIUS experiment can obtain positive
results.

\subsection{Triple maximal mixing scheme}

\begin{figure}[!t]
\hbox to \hsize{\hfil\epsfxsize=7cm\epsfbox{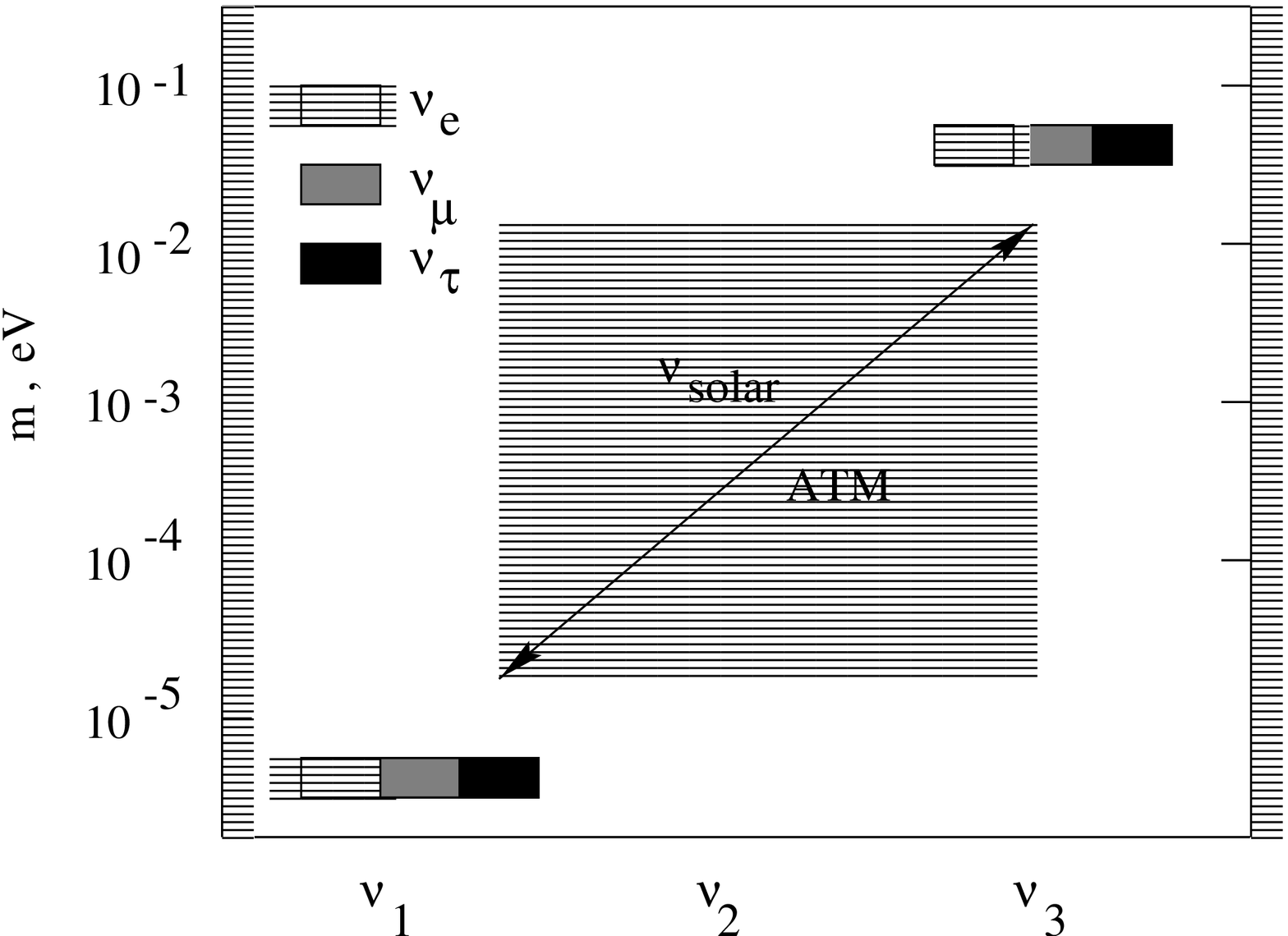}\hfil}
\caption{~~Neutrino masses and mixing  in  the  
scheme with threefold maximal mixing.
}
\label{ftriple}
\end{figure}

In the scheme of \cite{perkins} all elements of
the mixing matrix are assumed to be
equal: $|U_{ij}|=1/\sqrt{3}$ (see fig. \ref{ftriple}). The
\bbmass
is dominated by the contribution from the third state: 
\be
m_{ee}^{(3)}=\frac{1}{3}\sqrt{\Delta m^2_{atm}}~. 
\ee 
The best fit of the atmospheric neutrino data in this scheme 
implies that 
$\Delta m^2_{atm}\simeq 8 \cdot 10^{-4}~ {\rm eV}^2$ and thus
\be
m_{ee} \approx 10^{-2} {\rm eV}~.  
\ee 
The scheme has rather definite predictions 
for solar and atmospheric neutrinos. It does not give a 
good fit of the data  
and will be tested by forthcoming 
experiments.

\section{Schemes with partial degeneracy}

In the case of partial mass degeneracy,   
\be
\Delta m^2_{21} \ll m_1^2  \ll  \Delta m^2_{31}~, 
\label{ineq}
\ee
the
masses of the two light neutrinos are approximately equal to 
$m_1$ and the heaviest mass is determined by the 
atmospheric mass squared difference: 
\be             
m_1 \approx m_2, ~~~ m_3 \approx  
\sqrt{\Delta m^2_{31}} =  \sqrt{\Delta m_{atm}^2}~. 
\ee

The interval of masses implied by the condition of partial degeneracy 
(\ref{ineq}) is rather narrow
especially for the LMA and SMA  solutions of the solar neutrino problem, when
$\Delta m^2_{31}$ and $\Delta m^2_{21}$ differ by two orders of magnitude
only. A mass value of
$m_1 > 3 \cdot  10^{-2}$ eV will shift 
$m_3$  to larger values, and therefore 
influence the contribution from the third eigenstate.  
We will consider this ``transition" case separately in sect.
\ref{transregion}.

\begin{figure}[!t]
\epsfxsize=60mm
\centerline{\epsfbox{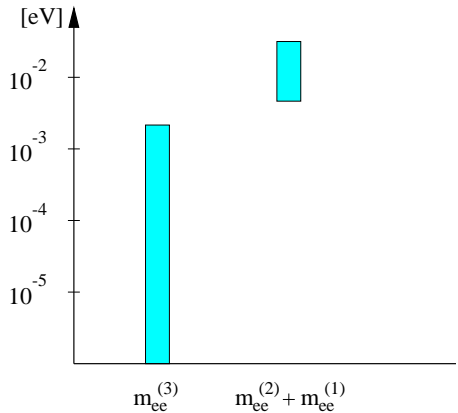}}
\caption{Contributions from different mass eigenstates to 
$m_{ee}$ for partially degenerate scenarios with MSW SMA solution.
\label{cstates5}}
\end{figure}

The contribution from the third state is the same as 
in hierarchical schemes (see fig. \ref{smix1}).
For the two light states, the contribution can be written as  
\be
m_{ee}^{(1)}+m_{ee}^{(2)}\simeq m_1 (\cos^2 \theta_{\odot} 
+ e^{i \phi_{2}} \sin^2 \theta_{\odot})~,  
\label{twolight}
\ee 
and depending on the relative phase $\phi_{2}$
it varies in the interval 
\be
m_{ee}^{(1)}+m_{ee}^{(2)} =  m_1 (\cos 2\theta_{\odot} - 1).  
\label{51}
\ee
This contribution can be further restricted,  if the
solution of the solar neutrino problem will be identified. 
In  the case of the SMA MSW  solution  
$m_{ee}^{(1)}$ dominates; the dependence on the phase practically
disappears and one gets 
\be
m_{ee}^{(1)} + m_{ee}^{(2)} \simeq m_1. 
\ee
The condition of partial degeneracy  
implies that  the mass $m_1$ should be 
in the interval:  
$$
0.5  \cdot 10^{-2} {\rm eV} <  m_1 < 3 \cdot 10^{-2} {\rm eV},   
$$  
and therefore,  $m_1$ can reach  $3 \cdot 10^{-2}$ eV at most 
(fig. \ref{cstates5}). 

Summing up all contributions we expect $m_{ee}$ between 
$10^{-3}$  and  $3 \cdot10^{-2}$  eV. 
Notice that a lower bound on $m_{ee}$ exists here. 
Near the upper bound the mass $m_{ee}$ is dominated by the contribution 
from the lightest states and therefore the \znbb decay rate will give a direct
measurement  of  $m_1$: $m_1 \approx m_{ee}$.

Observations of $m_{ee}$ 
larger than $m_{ee}^{(3)}= U_{e3}^2 m_3$ 
($m^{(3)}_{ee}$ can be determined from 
oscillation experiments) 
would favor the scheme,
although will not allow one to identify it unambigously.

Future observations 
of  \znbb decay with  $m_{ee} > 3 \cdot 10^{-2}$ eV will exclude the 
scheme testifying for spectra with complete degeneracy or inverse
hierarchy (see sect. 5 or 7).\\

For the LMA  solution the
typical $\Delta m^2_{21}$ is bigger than in the SMA case
and the condition of partial degeneracy 
implies an even narrower interval $m_1 = (1 - 3) \cdot 10^{-2} {\rm eV}$.  
Moreover, for $m_1$ at the lower limit of this interval,   
the difference of light mass
eigenvalues can give a substantial correction to formula (\ref{twolight}). 
In the lowest approximation of $\frac{\Delta m^2}{m_1^2}$ we get: 
\be
m_{ee}^{(1)}+m_{ee}^{(2)} \simeq m_1 
(\cos^2 \theta_{\odot}
+ e^{i \phi_{2}} \sin^2 \theta_{\odot}) + 
e^{i \phi_{2}} \frac{\Delta m^2_{\odot}}{2 m_1} \sin^2 \theta_{\odot}. 
\label{twocorr}
\ee 
The correction (last term in this equation) can be as big as  
$10^{-3}$ {\rm eV} and may turn out to be important when a cancellation of
$m_{ee}^{(1)}$ and $m_{ee}^{(2)}$  occurs. 

\begin{figure}[!t]
\epsfxsize=60mm
\centerline{\epsfbox{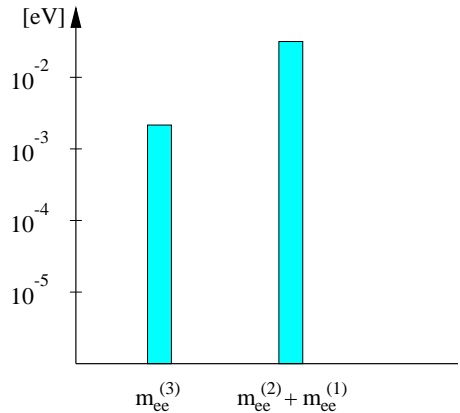}}
\caption{
Contributions to $m_{ee}$ from different mass eigenstates  
in schemes with partial degeneracy and  LMA, LOW or VO 
solutions of the solar neutrino problem. The 
degenerate states give the main contribution. A complete 
cancellation of contributions is  possible.
\label{cstates4}}
\end{figure}

Summing up the contributions we find, that the maximal value of 
$m_{ee}$ can be about $3 \cdot 10^{-2}$ eV as in the case of the
SMA solution with similar implications for future \znbb decay
searches. 
In contrast with the SMA  case, now due to possible strong cancellations 
of the 
contributions no lower bound on $m_{ee}$ can be obtained from the present
data (fig.\ref{cstates4}). 

Future oscillation results will allow to sharpen the predictions 
of $m_{ee}$. In particular, the solar neutrino experiments will allow to 
measure a  deviation  of mixing from the maximal value. 
The bound $1 - \sin^2 2\theta_{\odot} > 0.1$ would imply that 
$m_{ee}^{(1)} + m_{ee}^{(2)} > 3 \cdot 10^{-3}$ eV. In this case  no complete
cancellation in $m_{ee}$ is  possible and a minimum value 
$m_{ee}\geq 10^{-3}$ eV appears. The searches for $\nu_e$-oscillations 
driven by $\Delta m_{atm}^2$ will further restrict (or measure) 
$m_{ee}^{(3)}$.

Future studies of the  \znbb decay can have the following implications: 
(i) A measurement of $m_{ee} > 2 \cdot 10^{-2}$ eV will exclude the scheme. 
(ii) The non-observation of $m_{ee}$ at the level of $10^{-3}$ eV 
(second stage of GENIUS) can exclude the scheme if future oscillation
experiments will lead to a determination of  the sum 
$m_{ee}^{(1)} + m_{ee}^{(2)}$ and a lower bound on $m_{ee}$ will be derived. 
(iii) If $m_{ee}$ will be observed at the level $(0.3 - 2) 10^{-2}$ 
eV (and alternative schemes which yield a prediction in this interval 
will be rejected by other observations), then $m_{ee}$ measurements 
will imply a certain bound in the $m_1 - \phi_2$ plane.\\


In the case of the LOW solution 
$\Delta m^2_{\odot}$ is much smaller than for the LMA solution  
and $m_1$
can be in the
interval  $m_1 = (10^{-3} - 3 \cdot 10^{-2})$  eV. 
Correspondingly, the contribution from the two lightest states can be in 
the wider range $(10^{-4} -  3 \cdot 10^{-2})$ eV. 
The maximal value for $m_{ee}$ can reach $3 \cdot 10^{-2}$ eV. 
However it will be impossible to establish a lower bound on 
$m_{ee}$ even if the solar mixing angle $\theta_{\odot}$ will be measured.

Notice that the LOW solution can be identified by a 
specific enhancement of the regeneration effects (in particular the 
day/night asymmetry) in the lower energy part of the solar neutrino spectrum. 
An especially strong effect is expected on the $^7 Be$-line.

For vacuum oscillations the situation is similar to the LOW case.

\section{Schemes with  complete mass degeneracy 
\label{cdegerasy}}

In schemes with a degenerate neutrino mass spectrum  the
common mass $m_1$ is much larger than the mass splittings: 
\be 
\Delta m^2_{21} \ll  \Delta m^2_{31}  \ll m_1^2. 
\ee 
This can be realized  as long as $m_1>0.1$ eV.  

Already the present bound
$m_{ee} <  0.2 - 0.4$ eV \cite{hdmo} implies not too strong degeneracy unless 
a substantial cancellation of the contributions in $m_{ee}$ occurs.
Indeed,  if $\Delta m_{atm}^2 = 3 \cdot 10^{-3}$ eV$^2$, 
and $m_1 \sim 0.2$ eV, we get  
\be
\frac{\Delta m_{23}}{m_1} \approx \frac{\Delta m^2_{atm}}{2 m_2^2}= 
4 \cdot 10^{-2}.
\ee
For $m_1 = 0.1$ eV, the ratio equals 0.15. 

\begin{figure}[!t]
\epsfxsize=60mm
\centerline{\epsfbox{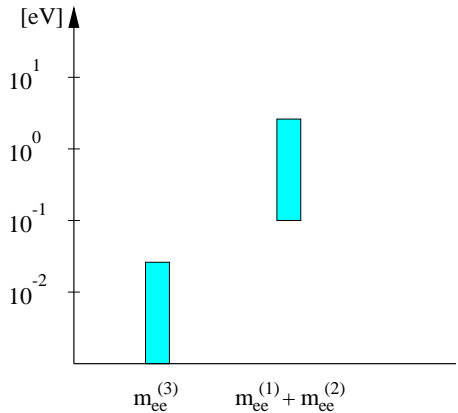}}
\caption{Contributions from different mass eigenstates to 
$m_{ee}$ for the scenario with  complete  mass degeneracy and  
the  SMA solution of the solar neutrino problem. 
\label{cstates8}}
\end{figure}

In the case of the SMA solution the $\nu_e$ flavor is mainly concentrated
in $\nu_1$ and 
\be
m_{ee}^{(1)}\simeq m_1 \cos^2 \theta_{\odot} \gg m_{ee}^{(2)} \gg m_{ee}^{(3)}.
\ee
Numerically, we get $m_{ee} \sim  m_{ee}^{(1)} \sim m_1 > 0.1$ eV,  
{\it i.e.} close to the present bound.  
Basically one measures $m_1$ by measuring the \bbmass 
(see fig.~\ref{cstates8}).

Important conclusions  follow from a  comparison 
of the \znbb decay results with the cosmological bounds 
on the neutrino mass \cite{bar99}, as well as from bounds which follow 
from observations of the large scale structure of the Universe. 
In this scenario one expects
\be
\sum m_i = 3 m_{ee}~, 
\ee
and therefore 
\be
\Omega_{\nu}=\frac{3 m_{ee}}{91.5 {\rm eV}}h^{-2}.
\label{omega}
\ee
Thus the effective Majorana mass, $\Omega_{\nu}$
and the Hubble constant are related. This relation may have 
the following implications:

1). A significant deviation from equation (\ref{omega})  
will exclude the scheme. 
The present bound on $m_{ee}$ implies $3 m_1 \leq (0.6 - 1)$
eV. If, {\it e.g.},   data on the large scale structure of the universe will
require $\sum m_i \simeq 1 $ eV, this scheme will be excluded
\cite{adh98,min97b}.

2). The discovery of  \znbb decay at the level of the present 
bound, 0.1 - 0.2 eV, will give  $\sum m_{i} \simeq 0.3-0.6$ eV.  
This range  can be probed by MAP and Planck. 
If these experiments will put a bound on the sum of neutrino masses 
below 0.3 eV the scheme will be excluded.

3). This scheme will also be excluded,  
if cosmological observations will require 
$\sum m_i > 0.3 $ eV,  but \znbb decay searches will give a bound below 
$m_{ee} < 0.1$ eV.

Apart from a confirmation of the SMA solution future oscillation experiments
will not influence predictions of $m_{ee}$ in this scheme. 

Observations of $m_{ee}$ at the level $0.1 - 0.4$ eV  will be in favor 
of the scheme.

\begin{figure}[!t]
\epsfxsize=10cm
\centerline{\epsfbox{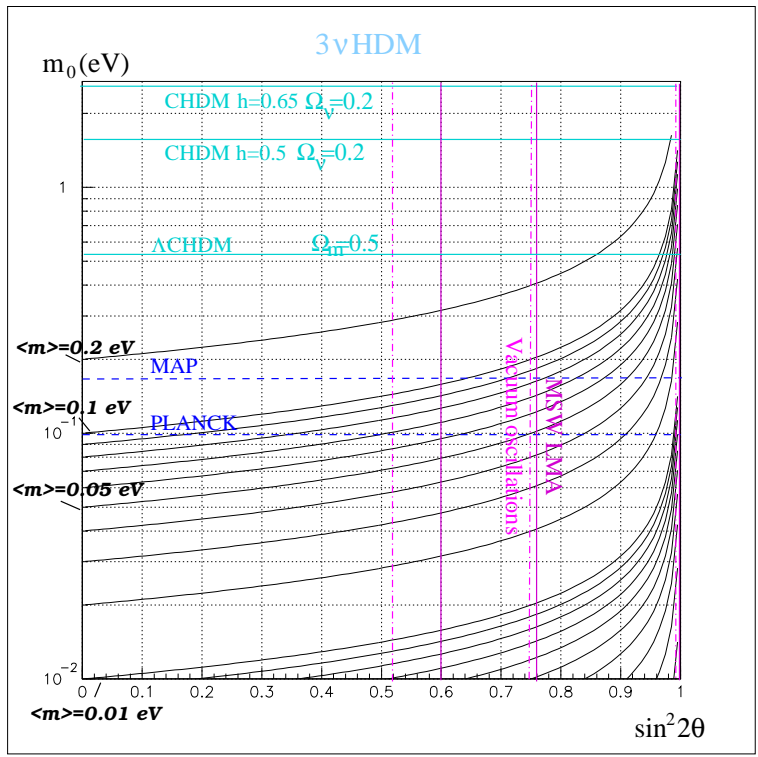}} 
\caption{
Plotted are iso-mass $\langle m \rangle = m_{ee}^{1+2}$
lines in the $m_1-\sin^2 2 \theta$ plane for 
the case of 
cancellation between the contributions $m_1$ and $m_2$ in degenerate 
scenarios. 
Mass splitting is neglected, since 
$m_2-m_1 \ll m_1$ and $m_3-m_1 < 0.1 m_1$ 
Shown are the bestfit values of $m_0$ for cold+hot dark matter (CHDM) 
for different values of the Hubble constant
(according to  \protect{\cite{eric98}})
and cosmological-constant+cold+hot dark matter
($\Lambda$CHDM) with $\Omega_m=0.5$ (according to 
\protect{\cite{Pri98}}) scenarios. 
Also shown is the sensitivity to $m_0$ of
MAP/Planck according to \protect{\cite{lopez,Eis98}}.
Vertical lines indicate 
the favored regions and best fits of the LMA and VO solutions 
according to \protect{\cite{Bah98}}.
\label{smix4}}
\end{figure}

\begin{figure}[!t]
\epsfxsize=60mm
\centerline{\epsfbox{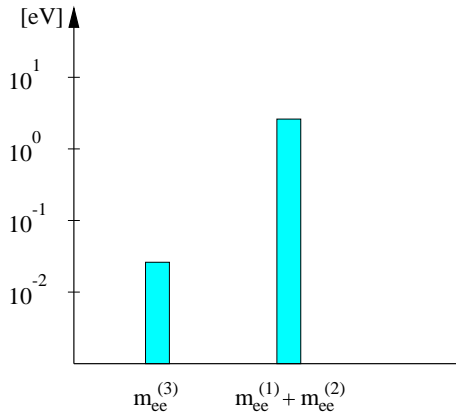}}
\caption{Contributions from different mass eigenstates to 
$m_{ee}$ for the scenario with complete degeneracy and  LMA,
LOW   or VO solution. 
\label{cstates9}}
\end{figure}

For the LMA solution a significant cancellation of the contributions from the
first and the second 
state may occur, resembling the situation in the partially degenerate case. 
We have
\be
m_{ee}^{(1)}+m_{ee}^{(2)} \simeq m_1 (\cos^2 \theta_{\odot}
+ e^{i \phi_{2}} \sin^2 \theta_{\odot})~, 
m_{ee}^{(3)} = m_1 |U_{e3}|^2~. 
\label{degen}
\ee
Since $U_{e3}^2\leq 0.03$,  
the contribution from the first two states dominates (fig.~\ref{cstates9}),    
unless strong mixing, which has an extremely small deviation from maximal 
mixing, is introduced: 
$|1 - \sin^2 2\theta_{\odot}| < 10^{-3}$, which leads to strong 
cancellation.  Thus
\be
m_{ee} \approx  m_{ee}^{(1)}  + m_{ee}^{(2)} = 
(\cos 2\theta_{\odot} - 1) \cdot m_1 = 
(0.2 - 1) m_1~, 
\ee
and since $m_1 > 0.1$ eV,  we expect for $\sin^2 2\theta =0.96$: 
\be
m_{ee}\geq 2 \cdot 10^{-2} {\rm eV}.
\ee 
Notice that some recent studies show that even exact maximal mixing 
is allowed by the present data,  so that the cancellation can be 
complete.

Precise measurements of $\theta_{\odot}$ in future oscillation
experiments will play a crucial role for predictions of 
the mass $m_{ee}$.  

If the scheme will be identified, measurements of $m_{ee}$ 
will provide a  bound in the $m_1 - \phi$-plane.

The same results hold also for the LOW solution. \\

For the vacuum oscillations of the $\nu_{\odot}$-problem, 
the situation is similar to the one with LMA
MSW. 
In the strict bi-maximal scheme  $U_{e3}^2=0$ and
$U_{e1}^2 = U_{e2}^2$,  so that for $\phi_2=\pi$
$m_{ee} \equiv 0$ in the limit of equal
masses. Small deviations from zero can be related 
to the mass difference of $m_1$  and $m_2$:
\be
m_{ee}\simeq \frac{1}{2}(m_1-m_2)= \frac{1}{4} 
\frac{\Delta m^2_{\odot}}{m_1}
\simeq 10^{-10} {\rm eV}\frac{1 {\rm eV}}{m_1}.
\ee
Thus,  no unique prediction for $m_{ee}$ exists. Although a large value 
$m_{ee}> 0.1$ eV would favor degenerate scenarios, the non-observation
of \znbb decay at the level of $0.1$ eV will not rule out the scheme.

The identification of the scenario will require
(i) a strong upper bound on $U_{e3}$, (ii) the confirmation of the 
vacuum oscillation solution 
and (iii) a large $m_{ee} > 0.1$ eV. This would testify
for the case of addition of the $\nu_1$ and $\nu_2$ contributions.
An upper bound on $m_1$ can be obtained from cosmology.

If however \znbb decay will not be discovered it will be practically impossible
to exclude the scenario (and  distinguish it from the hierarchical cases),
unless cosmology will be able to measure neutrino masses down to
$m_1 \simeq 0.1$ eV. 
The key element is the precise determination of $\theta_{\odot}$ 
and its deviation from the maximal mixing.\\

Let us consider how deviations from  the exact 
bi-maximal case 
affect the predictions for the rate in double beta decay experiments.
The lower bound on the effective mass turns out to be
\be 
m_{ee} \simeq m_1 \sqrt{1 - \sin^2 2\theta}~. 
\ee
We show this result 
in fig. \ref{smix4} as lines of minimal values of $m_{ee}$ 
in the $m_1 - \sin^2 2 \theta$
plane. These lines give the lower bound on values of $m_1$ and the upper
bound on values of 
$\sin^2 2\theta$ for which a given value $m_{ee}$ can be 
reproduced. 
We have shown also  the favored regions of the solar
MSW large mixing angle solution as well as the ``Just-so'' vacuum 
oscillation solution. 
E.g., a $\Lambda$CHDM model with a total $\Omega_m=0.5$ of
both cold and hot dark matter as well 
as a cosmological constant, and a Hubble constant of $h=0.6$ would imply an 
overall mass scale of about 0.5 eV.
Assuming a mixing corresponding to the 
best fit of solar  large mixing MSW or vacuum oscillations, 
$\sin^2 2 \theta=0.76$, this yields 
$\langle m \rangle=0.2-0.5$ eV. 
Larger mixing allows for smaller values of 
$m_{ee}$. In fig. \ref{smix4} also shown is the sensitivity of CMB studies
with MAP and Planck, which have been
estimated to be sensitive to $\sum m_{\nu}=0.5-0.25$ eV \cite{Eis98,lopez}.
For not too large mixing already the present \znbb decay
bound ontained from the Heidelberg-Moscow 
experiment \cite{hdmo} is close to the sensitivity of these cosmological
observations.

\section{Transition regions 
\label{transregion}
}

There are two intermediate regions of $m_1$: 

1) The region with $m_1 \sim \sqrt{\Delta m^2_{\odot}}$,  where the
transition between the hierarchical case and the case with 
partial degeneracy occurs. Here $m_1 \sim m_2 \ll m_3$. 

2). The region with $m_1 \sim \sqrt{\Delta m^2_{atm}}$ which corresponds 
to the transition between partial degeneracy to complete 
degeneracy: $m_1 \approx m_2 \sim m_3$.
Here the two lightest states are strongly degenerate 
and their contributions are described by eq. (\ref{51}) 
in sect. 4.     
with the only difference that $m_1$ now can be larger.  
For $m_1^2 \sim  \Delta m^2_{atm}$,  $m_1$ and therefore 
$m_{ee}$ can reach 0.1 eV. 

The contribution from the third state is also modified: 
\be
m_{ee}^{(3)} = |U_{e3}|^2 \sqrt{\Delta m^2_{atm}} 
\left[1 + \frac{m_1^2}{\Delta m^2_{atm}}\right]^{1/2}.  
\label{thirdcor}
\ee
Thus now for the same values of the oscillation parameters 
the contribution $m_{ee}^{(3)}$ can be $[1 + m_1^2/\Delta
m^2_{atm}]^{1/2}$ 
$\sim (1 - 2)$ times larger. 

\begin{figure}[!hhh]
\parbox{5cm}{
\epsfxsize=120mm
\epsfbox{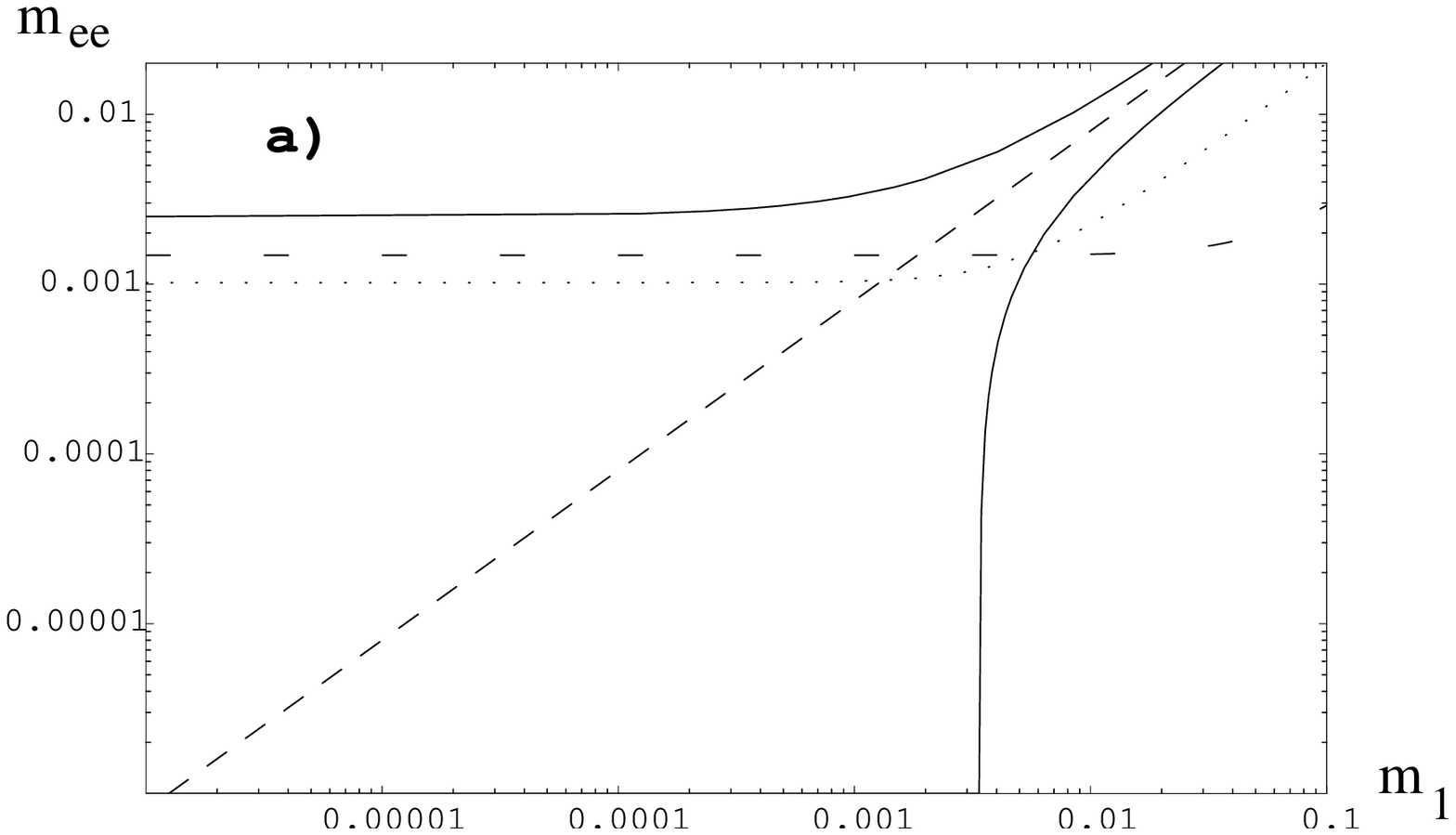}
}
\parbox{5cm}{
\epsfxsize=120mm
\epsfbox{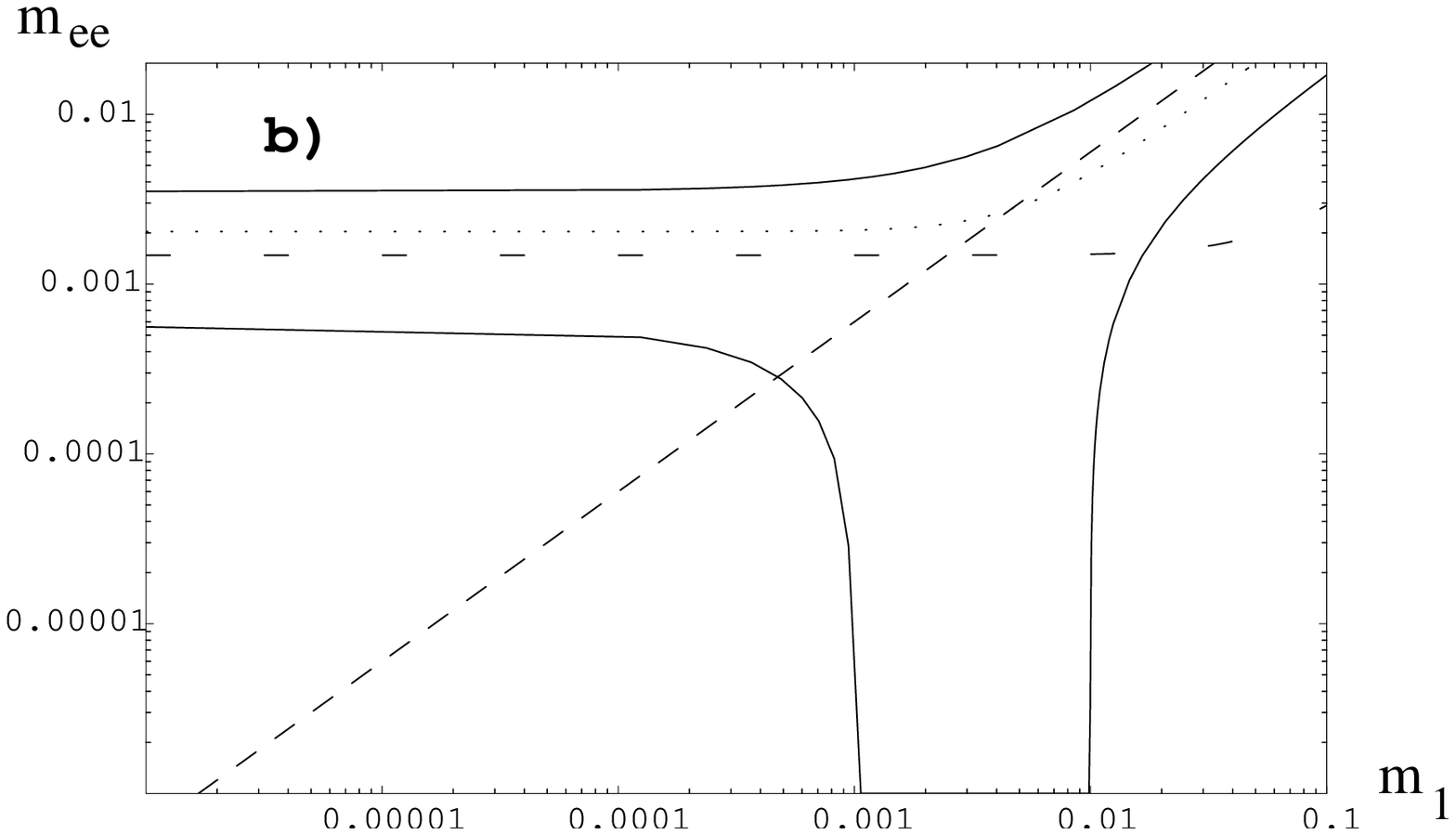}
}
\parbox{5cm}{
\vspace*{-4cm}
\epsfxsize=120mm
\epsfbox{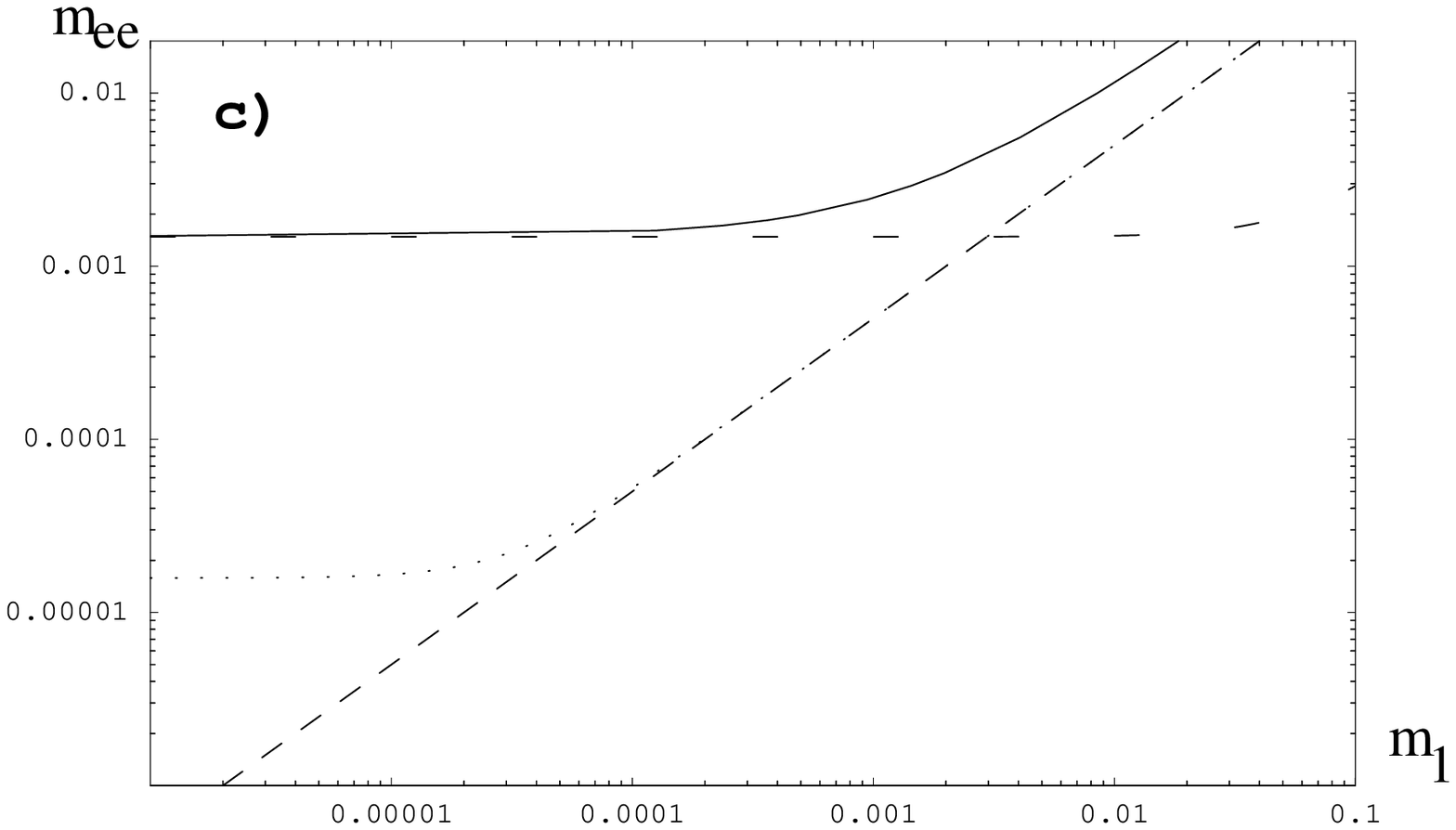}
}
\vspace*{-6cm}
\\
\\
\\
\\
\end{figure}

\begin{figure}
\parbox{10cm}{
\vspace*{-1cm}
\epsfxsize=120mm
\epsfbox{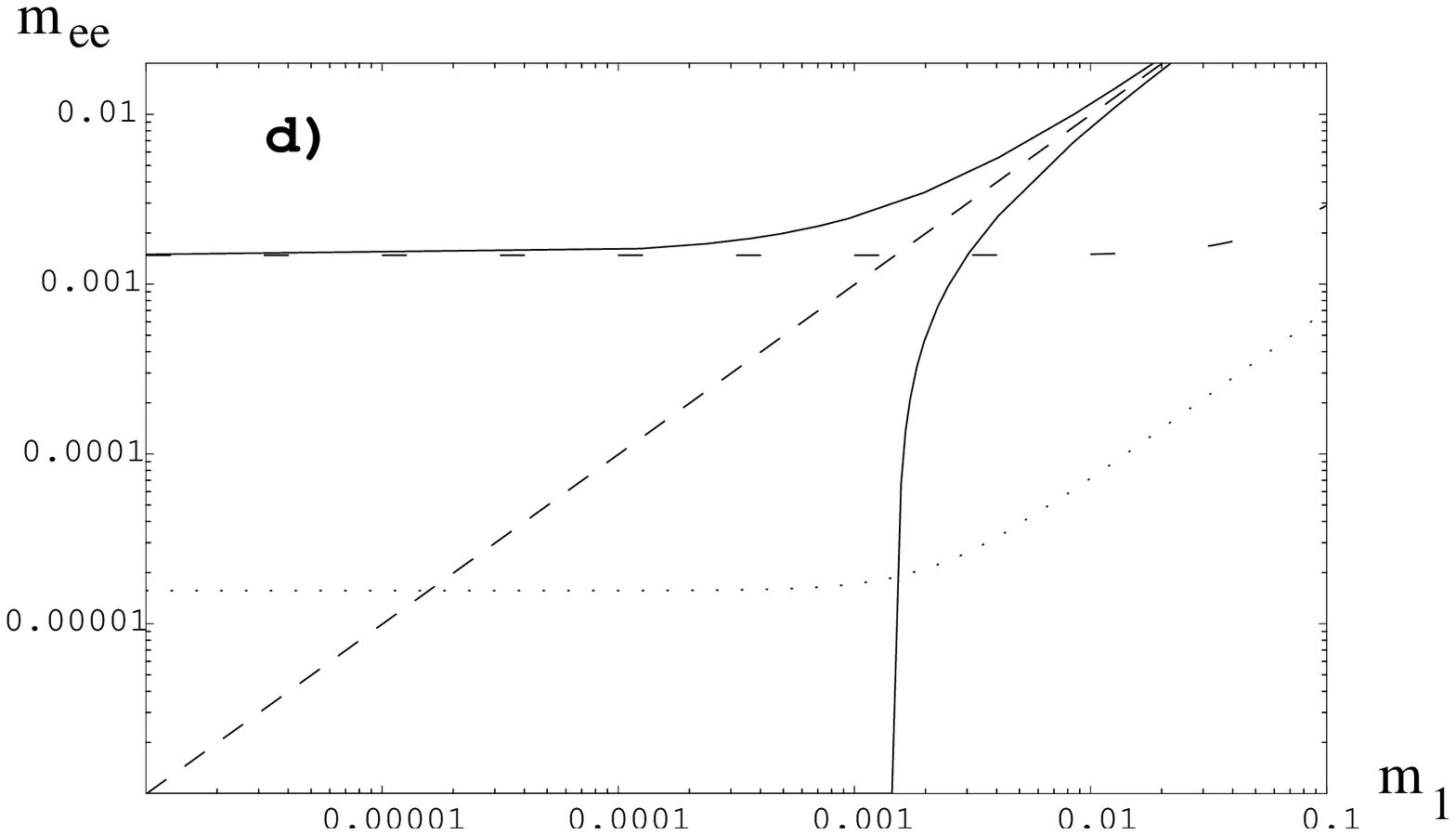}
\vspace*{-2cm}
}
\parbox{10cm}{
\vspace*{-1.5cm}
\epsfxsize=120mm
\epsfbox{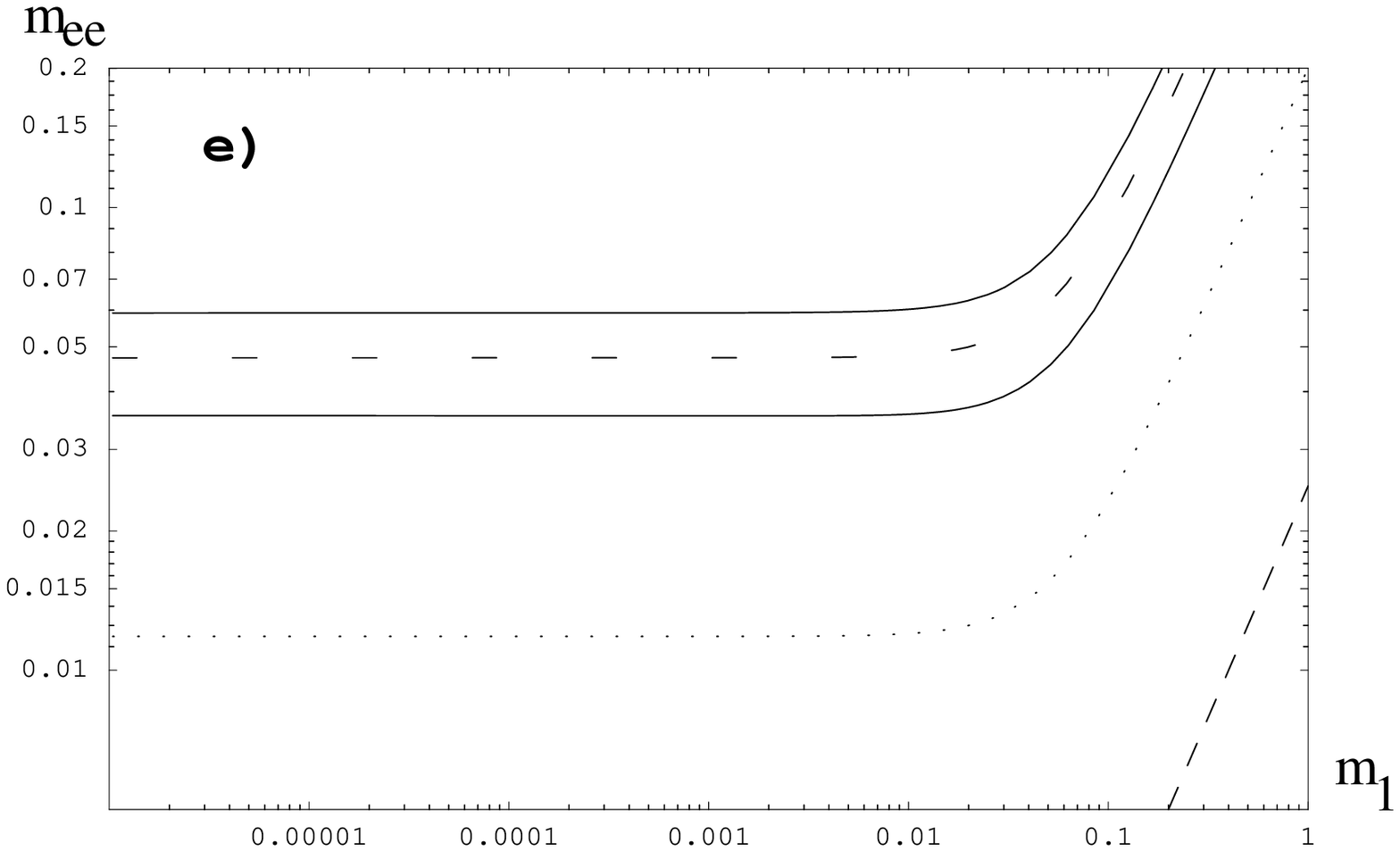}
\vspace*{-2cm}
}
\parbox{10cm}{
\vspace*{-2cm}
\epsfxsize=120mm
\epsfbox{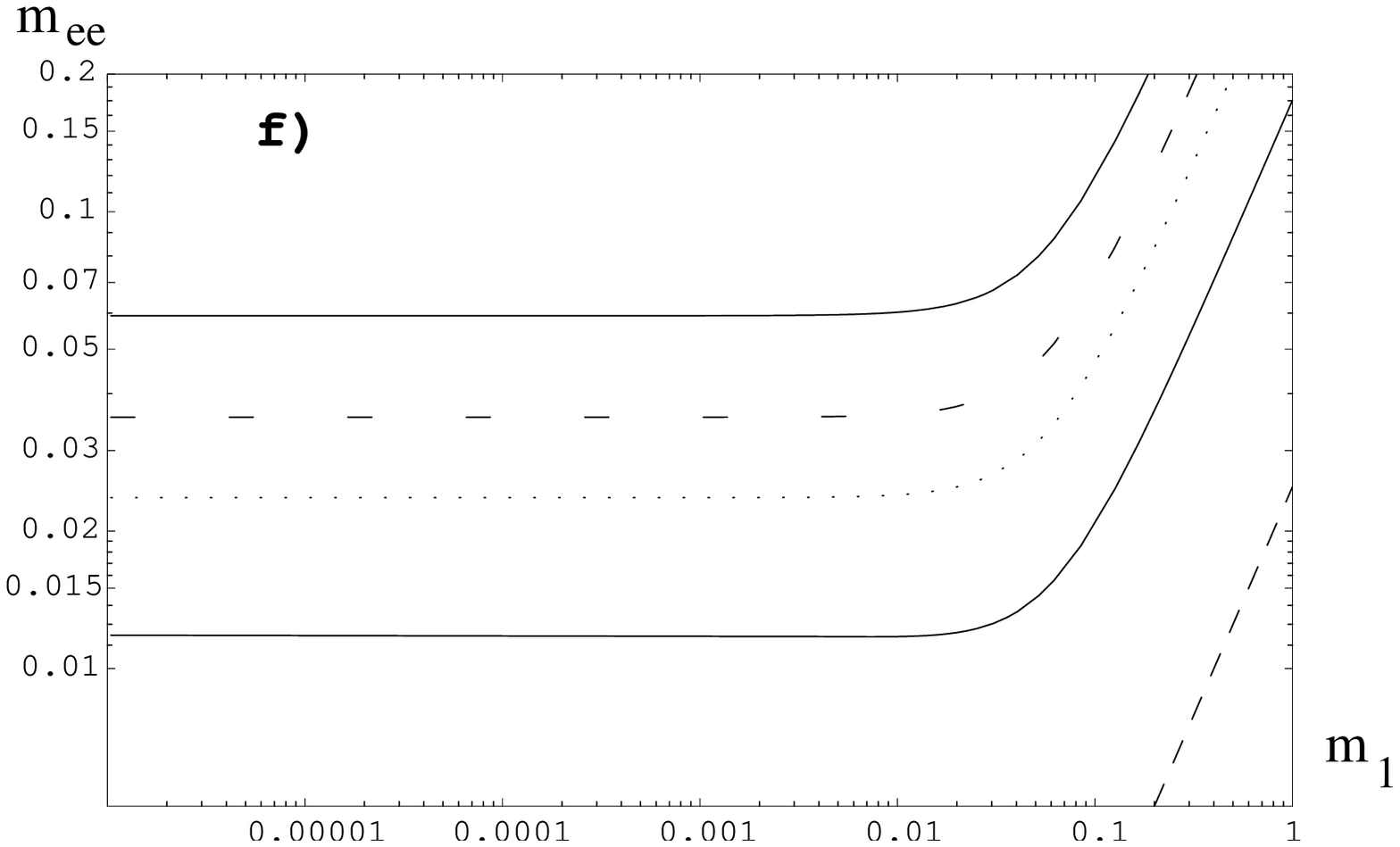}
}
\\
\end{figure}

\begin{figure}[!th]
\parbox{10cm}{
\epsfxsize=120mm
\vspace*{-3cm}
\epsfbox{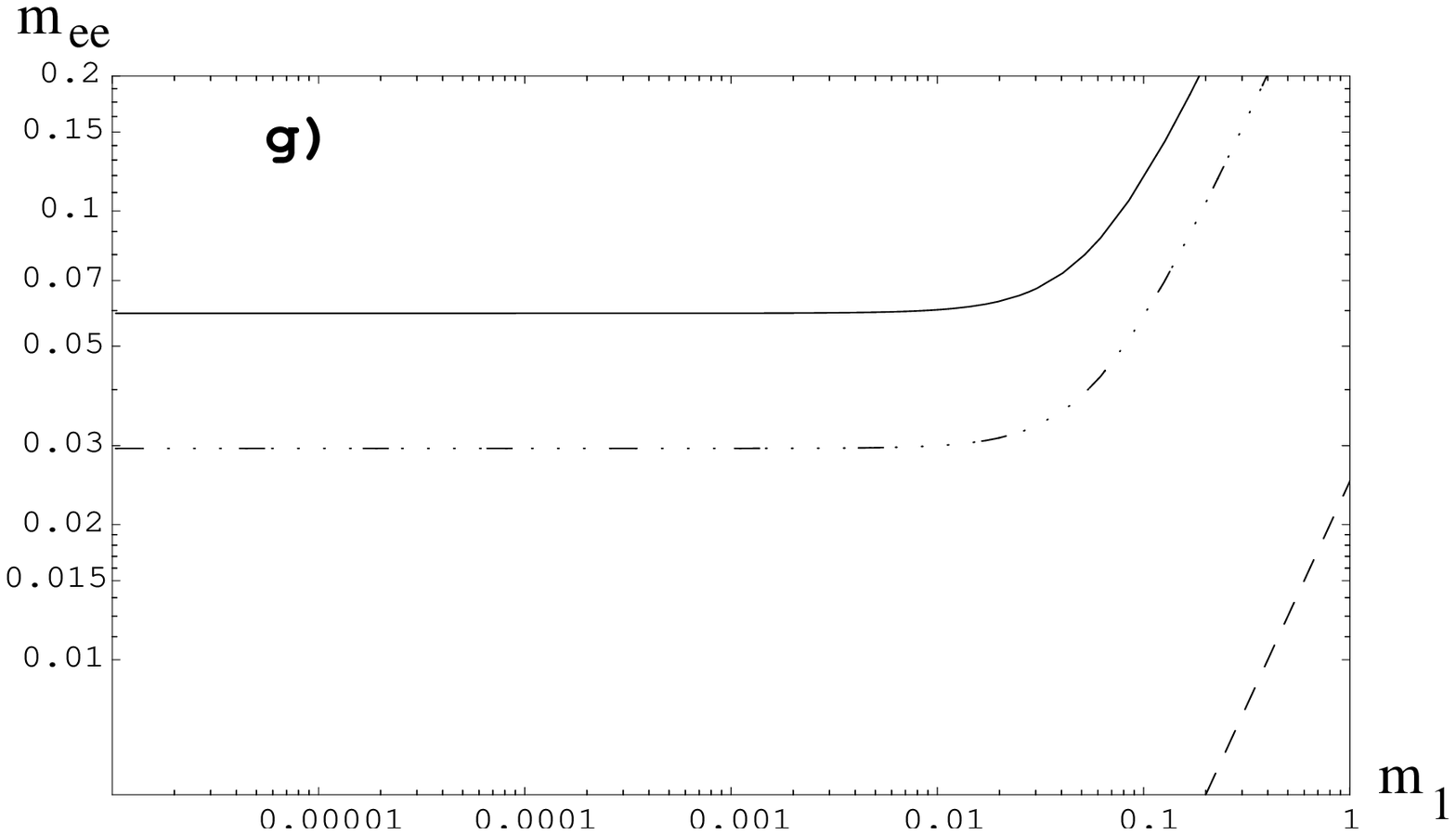}
}
\parbox{10cm}{
\epsfxsize=120mm
\epsfbox{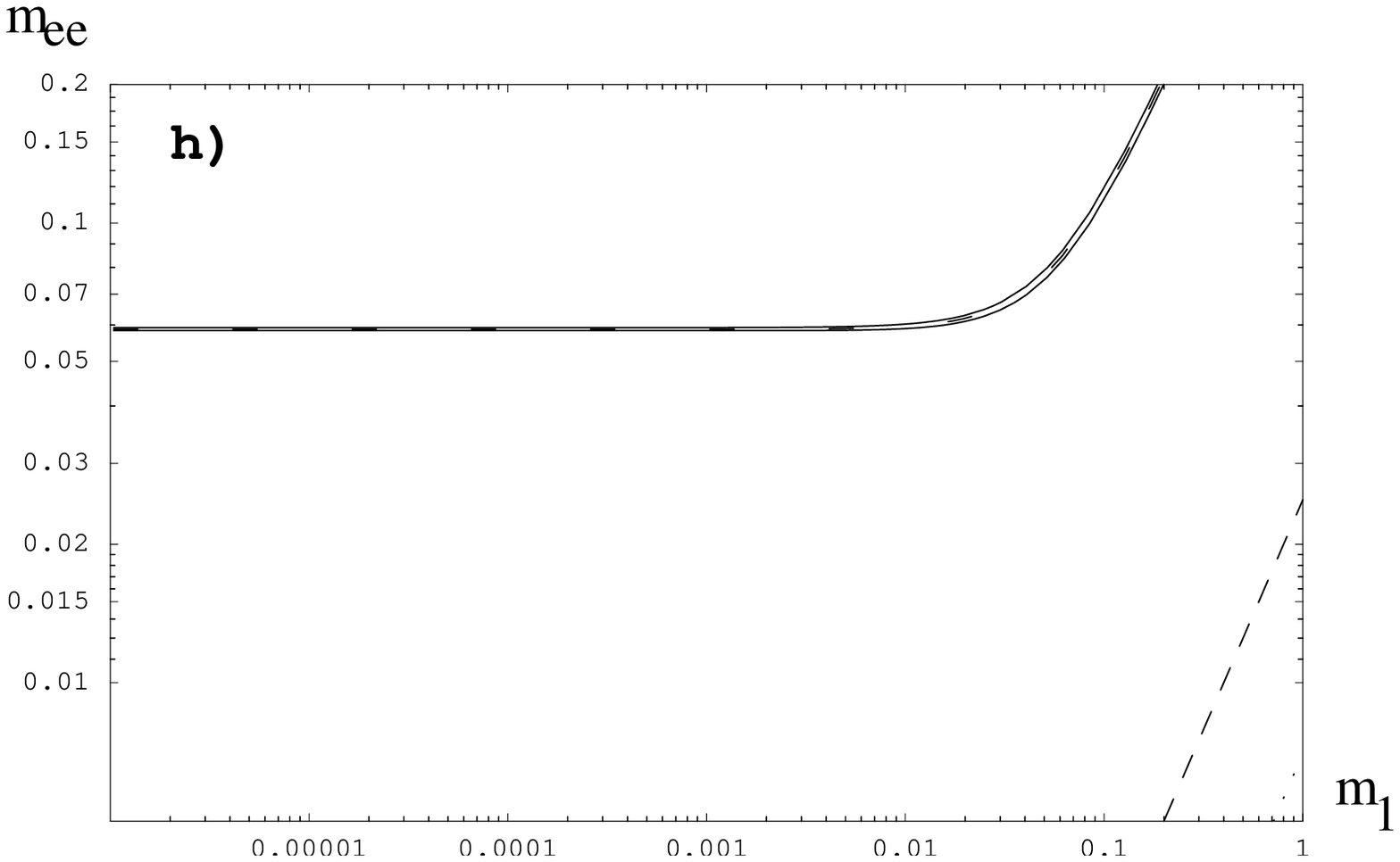}
}
\caption{
$m_{ee}$ (eV) as a function of $m_1$ (eV) for three-neutrino mixing.
Shown are the contributions $m_{ee}^{(1)}$ (dashed), $m_{ee}^{(2)}$
(dotted) and $m_{ee}^{(3)}$ (interrupted dashes).
The solid lines correspond to $m_{ee}^{max}$ and $m_{ee}^{min}$ and show 
the allowed region for $m_{ee}$.
Panels a)-d) correspond to the case for normal hierarchy, panels e)-h) 
-- inverse hierarchy. Shown are the cases a) and e) 
LMA MSW with $U_{e2}^2=0.2$, b) and f) LMA MSW with $U_{e2}^2=0.4$,
c) and g) vacuum oscillation with $U_{e2}^2=0.5$ and d) and h) SMA MSW with 
$U_{e2}^2=7\cdot10^{-3}$. 
The mixing of
the third state is varied from zero to its upper bound,
$U_{e3}^2=2.5 \cdot 10^{-2}$.
\label{algen}}
\end{figure}

In fig.\ref{algen} we show the dependence of the individual
contributions  to $m_{ee}$ on $m_1$.  
For $m_{ee}^{(3)}$ only the upper bound is used; the two other 
lines represent possible values of $m_{ee}^{(1)}$ and $m_{ee}^{(2)}$ 
for certain 
neutrino mixing parameters.  
We show also the maximal and the minimal possible values of $m_{ee}$.  

The position of the first transition region is determined by the
specific solution 
of the solar neutrino problem: According to fig. \ref{algen}, 
$m_1 = (2 - 15)\cdot 10^{-3}$ eV for the LMA solution, 
$m_1 = (1 - 9)\cdot 10^{-3}$ eV for the SMA MSW solution 
and $m_1 = (1 - 10)\cdot 10^{-5}$ eV for the VO solution.

The position of the second transition region, 
$m_1 = (3 - 20)\cdot 10^{-2}$ eV, is similar in all the cases. 

The upper bounds on $m_{ee}$  as functions of $m_1$ have a similar
dependence for all the cases. The lower bounds are different  
and depend on specific values of oscillation parameters.

Thus,  for the LMA solutions 
a lower bound exists  in the range of mass hierarchy 
($m_1  < 10^{-3}$ eV) if the solar mixing angle is sufficiently 
large  (see fig. \ref{algen} b)). In this case the contribution 
from $\nu_2$ dominates and no cancellation is possible 
even for maximal possible $m^{(3)}_{ee}$. In contrast, 
for a lower $\sin^2 2\theta_{\odot}$ the cancellation can be
complete so that no lower bound appears (see fig. \ref{algen} a)). 

In the first transition region all states contribute with comparable portions 
to $m_{ee}$, thus cancellation is possible and no lower bound exists.

In the second transition region as well as in the completely 
degenerate case the first and the second state give the dominating
contributions to $m_{ee}$ and the increase of $m_3$ 
does not influence significantly the total $m_{ee}$. 
The mass $m_{ee}$ is determined by $m_1$ and $\theta_{\odot}$. 
Moreover, a larger  $\sin^2 2\theta_{\odot}$ implies a larger  
possible range of $m_{ee}$ for a given $m_1$ (fig. \ref{algen} a,b)).

Let us consider the SMA MSW solution (fig. \ref{algen} d)).  
In the mass hierarchy region the third 
state gives the main contribution and no lower bound exists. 
A lower bound on $m_{ee}$ appears at $m_1 > 1.5 \cdot 10^{-3}$ eV 
and at $m_1 > 10^{-2}$ eV the mass $m_{ee}$ is given by $m_1$.

In the case of the VO solution (fig. \ref{algen} c) 
the upper bound on $m_{ee}$ 
is given by $m^{(3)}_{ee}$ up to $m_1 \sim 2 \cdot  10^{-4}$ eV. 
In the range of partial degeneracy the contribution from the first and
the second states become important. No lower bound on $m^{(3)}_{ee}$
can be established from the present data in the whole range of $m_1$.

\section{Scheme with inverse mass hierarchy 
\label{invhier}} 

\begin{figure}[!t]
\hbox to \hsize{\hfil\epsfxsize=7cm\epsfbox{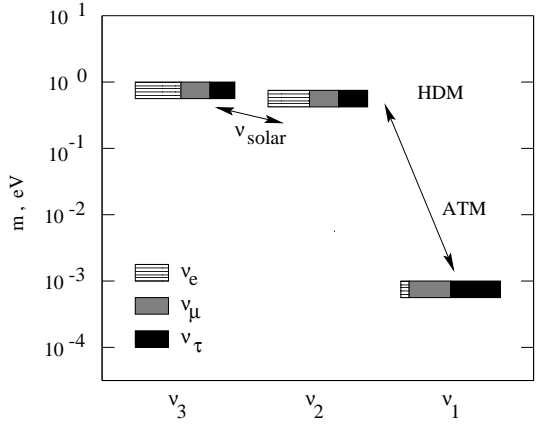}\hfil}
\caption{~~The neutrino mass and mixing pattern  in  the  bi-large
mixing  scheme with inverse  mass hierarchy.
}
\label{fbimaxinv}
\end{figure}

Let us consider the  partially degenerate spectrum with 
\be
m_3^2 \approx m_2^2 = \Delta m_{atm}^2, ~~~ 
m_1^2 \ll  m_2^2, ~~~ \Delta m_{23}^2 = \Delta m_{\odot}^2~, 
\label{degen2}
\ee
so that the mass of the second and third
neutrino  are determined from the    
atmospheric neutrino data. 
The $\nu_e$ flavor is concentrated in the heavy states 
(inverse mass hierarchy). 
A small admixture of $\nu_e$ in the 
lightest state can exist (fig. \ref{fbimaxinv}  ).

The contribution to $m_{ee}$ from the first state equals  
\be
m_{ee}^{(1)}=m_1 U_{e1}^2.   
\ee
The inequality  $m_1^2 \ll \Delta m_{atm}^2$ 
implies $m_1 < 2 \cdot 10^{-2}$ eV for $m_1^2/m_2^2 < 0.1$. 
Using then the CHOOZ result which restricts (in schemes with inverse 
hierarchy)  $U_{e1}^2$: 
$U_{e1}^2 < 2.5 \cdot 10^{-2}$,  we get 
\be
m^{(1)}_{ee} < 5 \cdot 10^{-4} ~ {\rm eV}~. 
\ee

\begin{figure}[!t]
\epsfxsize=60mm
\centerline{\epsfbox{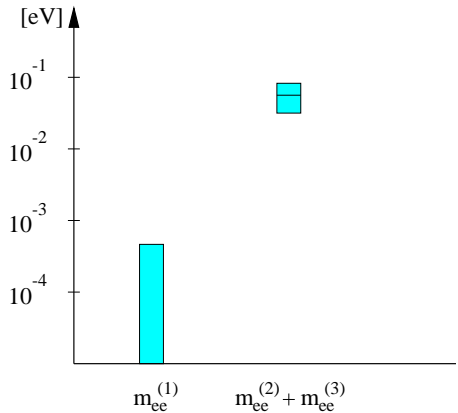}}
\caption{
Contributions to $m_{ee}$ 
from different mass eigenstates  
in the scheme with inverse mass hierarchy and  SMA solution.
The degenerate states give the main contribution, implying a unique
prediction for $m_{ee}.$
\label{cstates6}}
\end{figure}

The sum of the contributions from the two heavy degenerate states can be
written as  
\be
m_{ee}^{(2)}+ m_{ee}^{(3)} \simeq \sqrt{\Delta m^2_{atm}}(\sin^2
\theta_{\odot}
+ e^{i \phi_{23}} \cos^2 \theta_{\odot})~,  
\label{twothree}
\ee
where $\phi_{23} \equiv  \phi_{2} - \phi_{3}$. 
For the SMA solution we get from eq. 
(\ref{twothree}) 
\be
m_{ee} \approx m_{ee}^{(2)}+ m_{ee}^{(3)} \approx 
\sqrt{\Delta m^2_{atm}}= (4 - 8) \cdot 10^{-2} {\rm eV}
\label{mss4}
\ee 
and in the  bestfit point of the atmospheric neutrino data:   
$
m_{ee} \approx 6 \cdot 10^{-2} {\rm eV}. 
$
This means, that the predicted value  of  $m_{ee}^2$   
coincides with  $\Delta m^2_{atm}$ (fig.~\ref{cstates6}). 
This  coincidence provides a unique possibility to identify the scheme
(see also, e.g. \cite{pet99}).

The relation $m_{ee}^2 = \Delta m^2_{atm}$ 
applies also for the case of the LMA solution as long as $\phi_{2}=0$
\footnote{Notice that if the mass degeneracy originates from some  flavor
blind interactions one may indeed expect that the masses of 
$\nu_2$ and $\nu_3$ have the same phase.}.

For the LMA solution the sum of the contributions  
from the two heavy states lies in the interval 
\be
m_{ee}^{(2)} + m_{ee}^{(3)} = 
(\cos 2 \theta_{\odot}- 1)\cdot \sqrt{\Delta m^2_{atm}}.
\ee
For $\sin^2 2\theta_{\odot} < 0.98$ we get $m_{ee} > 4 \cdot 10^{-3}$ 
eV which is still much larger than $m^{(1)}_{ee}$. 
The compensation can be  complete if the mixing is  maximal. 
The  value  $m_{ee}^{(2)} + m_{ee}^{(3)} < 2 \cdot 10^{-3}$ eV 
requires a very small deviation from maximal mixing: 
$1 - \sin^2 2\theta_{\odot} <  2 \cdot 10^{-3}$.   
Thus, the lower bound on $m_{ee}$ can be further strengthened,  
if the deviation from  maximal mixing will be established. 

\begin{figure}[!t]
\epsfxsize=60mm
\centerline{\epsfbox{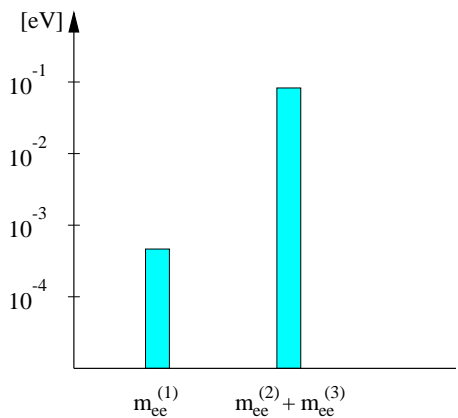}}
\caption{~~Contributions to $m_{ee}$ from different mass eigenstates  
in the schemes with inverse hierarchy and  
LMA, LOW or VO solutions.
Now cancellation between the degenerate states is possible leading to a 
wide range of values allowed for $m_{ee}$.}
\label{cstates7}
\end{figure}

A similar consideration holds for the cases of LOW MSW   
or vacuum oscillation solutions (see fig.~\ref{cstates7}). \\

The contribution of the two heavier eigenstates to the HDM,
$\Omega_{\nu} =  (2 m_1)$ 
$/(91.5~{\rm eV} h^2) \sim 0.01$, 
is rather small and below the reach of future projects on  
measurements of cosmological parameters. 

If the $\nu_e$ admixture in the lightest state is non-zero,  
so that the $\nu_e$-oscillations driven by $\Delta m^2$ 
exist, the scheme  can be identified by studying matter effects 
in atmospheric and  supernova neutrinos as well as in the long-baseline
experiments.

Indeed, in the case of inverse mass hierarchy the  
$\nu_e - \nu_3'$ level crossing (in matter) 
occurs in the {\it antineutrino} channel, so that
in supernovae  the antineutrinos $\bar{\nu}_e$ will be strongly converted
into a combination of $\bar{\nu}_{\mu}$, $\bar{\nu}_{\tau}$ and vice versa.
This leads to a hard $\bar{\nu}_e$'s 
spectrum at the Earth detector which coincides with 
the original $\bar{\nu}_{\mu}$ spectrum \cite{digh99}.

In atmospheric neutrinos the identification 
of the type of mass hierarchy will be possible 
if the sensitivity will be enough to 
detect  oscillation effects in  e-like events
(electron neutrinos and antineutrinos).
It will be also important to 
measure the sign of the electric charge of the lepton, 
since the matter effects are different 
in the neutrino and antineutrino channels 
and this difference depends on the 
type of mass hierarchy.

These matter effects 
can  be  studied in  LBL experiments \cite{Zub98,minos}
with neutrinos  
from  neutrino factories 
where beams of neutrinos and antineutrinos are well controlled.

Let us consider the dependence of the predictions for $m_{ee}$ on  
$m_1$. In the schemes with inverse hierarchy there is only one 
transition region: $m_1 \sim \sqrt{\Delta m^2_{atm}}$, that is 
$m_1 \sim  m_2 \approx m_3$ or $m_1 = (1 - 8)\cdot  10^{-2}$ eV. 
The sum of the contributions from the second and the third states 
dominates in the whole range of $m_1$. It is determined by the ``solar"  
mixing angle $\theta_{\odot}$ and $m_2 \approx m_3$. The latter 
changes from $\sqrt{\Delta m^2_{atm}}$ in the hierarchical region to 
$m_1$ in the region of complete degeneracy (see fig. \ref{algen} e-h).  
The mass  $m_{ee}$ is completely predicted in terms of $m_2$ for the SMA
solution (fig. \ref{algen} h).  

No lower bound on $m_{ee}$ appears when the (solar) mixing 
parameter is maximal or close to maximal. \\

\section{Four  neutrino scenarios}
\label{fournu}

The
introduction of new (``sterile") neutrinos mixed with the usual 
SU(2) doublet neutrinos opens 
new possibilities for the construction of the neutrino mass spectrum 
and for the explanation of the data. 
It also modifies predictions of $m_{ee}$.  
Here we will consider several 
scenarios which are  motivated both by  phenomenology
and theory. 
All scenarios we will discuss contain one or two (degenerate) 
states in the  range relevant for structure formation in the
universe and/or for the LSND oscillations.

\subsection{Scenario with small flavor mixing and mass hierarchy}

The scheme (fig. 20) is characterized  by a mass hierarchy: 
\be 
m_4 = m_{HDM}, ~~ m_3 \approx \sqrt{\Delta m_{ATM}^2}, ~ 
m_2 \approx \sqrt{\Delta m_{\odot}^2}, ~~ m_1 \ll m_2 ~. 
\ee
The states $\nu_{\mu}$ and $\nu_s$ are strongly mixed in 
the second and fourth mass eigenstates, so that 
$\nu_{\mu} \leftrightarrow \nu_s$ oscillations 
solve the atmospheric neutrino problem.  
All other mixings are small. In particular, 
the solar neutrino problem is  solved by small mixing 
MSW conversion $\nu_{e} \rightarrow  \nu_{\mu}, \nu_s$. 

The main motivation for this scheme is to 
avoid the introduction of large mixing between flavor states 
and to keep in this way as much as possible correspondence with the quark
sector. 

A clear signature of the scheme is the $\nu_{\mu} \leftrightarrow  \nu_s$
oscillation solution of the atmospheric neutrino problem.  
The solution  can be tested by 
(i) studies of the neutral current interactions in 
atmospheric neutrinos, in particular,  $\nu N \rightarrow \nu N \pi^0$ 
(with $N=n,p$), 
which gives the main contribution to the sample of the 
so called $\pi^0$ events  
(the rate should be  lower in the $\nu_{\mu} \rightarrow  \nu_s$ 
case); 
(ii) studies of the zenith angle distribution of the 
upward going muons (stopping and through-going);  
(iii) detection of the $\tau$ leptons produced by converted 
$\nu_{\tau}$.  

Recent Super-Kamiokande data  do not show a deficit of 
$\pi^0$ events, and moreover the $\nu_{\mu} \rightarrow  \nu_{\tau}$
oscillations give a  better fit  (of about $2-3 \sigma$)
of the zenith angle distribution thus favoring the 
$\nu_{\mu} \rightarrow  \nu_{\tau}$ interpretation. 
However, more data are needed to draw a definite conclusion 
(see \cite{gonz2000}).  

\begin{figure}[!t]
\hbox to \hsize{\hfil\epsfxsize=8cm\epsfbox{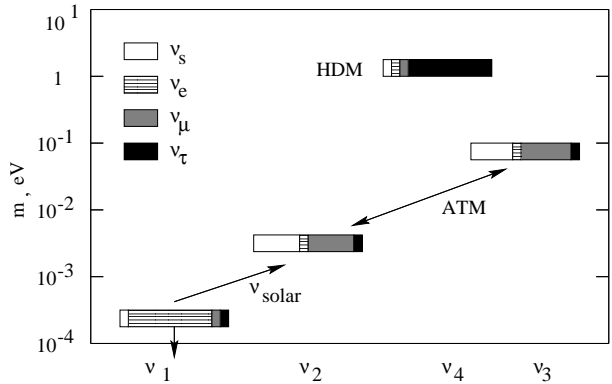}\hfil}
\caption{~~Neutrino masses and mixing in the
4 $\nu$ scenario with small flavor mixing and mass hierarchy. Here the white
parts of the boxes correspond to admixtures of the sterile state.}
\label{fint}
\end{figure}

The novel element of this scheme (compared with the 
3$\nu$ - schemes discussed in the previous sections) is the existence of a 
heavy state in the HDM range. Its contribution to 
$m_{ee}$ equals: 
\be
m_{ee}^{(4)} = U_{e4}^2 m_4~.
\ee
The relevant parameters,  
$U_{e4}$ and $m_4$,  can be determined 
from studies of the short range 
$\nu_e \leftrightarrow  \nu_e$ oscillations (disappearance)   
driven by the largest mass splitting  
$m_4 \approx \sqrt{\Delta m^2} \approx m_{HDM}$. For  this channel the
effective mixing angle equals  
\be 
\sin^2 2\theta_{ee} = 4 |U_{e4}|^2(1 -  |U_{e4}|^2)~ \approx 4|U_{e4}|^2,  
\ee
so that 
\be
m_{ee}^{(4)} \approx \frac{1}{4}\sqrt{\Delta m^2} \sin^2 2\theta_{ee}. 
\ee 
The corresponding iso-mass lines in the $\Delta m^2 -\sin^2 2\theta$
plot together with various oscillation bounds are shown in fig. 21. 

\begin{figure}[!t] 
\epsfxsize=120mm
\epsfbox{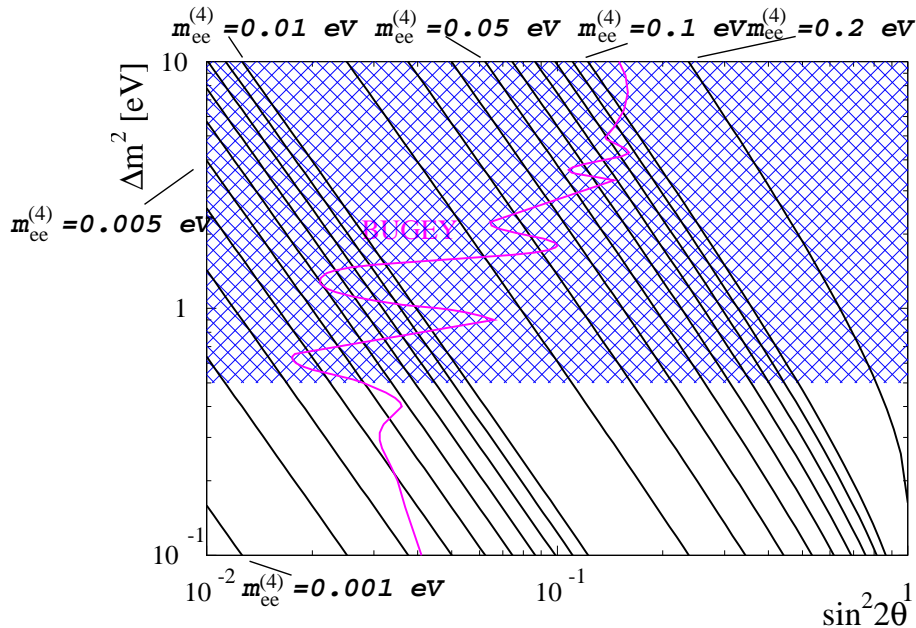}   
\caption{Iso-mass $|m_{ee}^{(4)}|$ lines in the four-neutrino scenario
with small flavor mixing and mass hierarchy. 
The shadowed area shows the region for neutrino masses of cosmological
interest  as HDM. Also shown are the regions excluded by the reactor
experiment BUGEY  (from \protect{\cite{lella}}).} 
\label{iso4nu} 
\end{figure}

In the cosmologically interesting range, 
$\sqrt{\Delta m^2} \approx m_{HDM} \simeq (0.5 - 5)$ eV, 
the mixing is constrained by 
the BUGEY experiment: $\sin^2 2 \theta_{ee}= (2 - 4) \cdot
10^{-2}$. Therefore we get the upper bound   
\be
m_{ee}^{(4)} \simeq (2 - 5) \cdot 10^{-2} {\rm eV}~. 
\ee


There is no strict relation between $m_{ee}$
and the parameters of the $\nu_{e} \leftrightarrow \nu_{\mu}$
oscillations since both relevant mixing elements 
$U_{e3}$ and $U_{\mu 3}$ are small. Indeed, now the effective 
depth of oscillations is determined by 
$\sin^2 2\theta_{e\mu} = 4 |U_{e3}|^2 |U_{\mu3}|^2$, so that 
\be 
m_{ee}^{(4)} = \frac{\sin^2 2\theta_{e\mu} m_{HDM}}{4 |U_{\mu4}|^2}~. 
\label{info}
\ee 
It is impossible to infer useful information from this 
unless the $U_{\mu4}$ will be determined from other 
experiments. Taking the  bound $|U_{\mu4}|^2 < 0.25$ 
from the  $3 \nu$ -  analysis of the atmospheric neutrino data,   
we get from eq. (\ref{info}) the lower bound  
\be
m_{ee}^{(4)} >  \sin^2 2\theta_{e\mu} m_{HDM}~. 
\ee

To get an estimation we assume  
$\sin^2 2 \theta_{e \mu} \sim 10^{-3}$ 
which corresponds to upper bounds on the elements 
$U_{e4}$ and  $U_{\mu4}$ from the BUGEY experiment and 
searches for $\nu_{\mu} \leftrightarrow \nu_{\tau}$ oscillations. 
(Notice that  LSND result can not be completely explained in this scheme.) 
This leads to 
\be
m_{ee}^{(4)} \gsim  10^{-3} {\rm eV}~.
\ee

The contributions from the three light states are similar  
to the  contributions in the 
$3\nu$ single maximal mixing scheme with mass hierarchy (sect. 3.1).
In particular, the largest contribution may come  from the third mass
eigenstate:  
$m_{ee}^{(3)} = \sqrt{\Delta m^2_{atm}} U_{e 3}^2 < 2 \cdot 10^{-3}$ eV  
(see eq. (\ref{thirdmax})). The contributions from the two 
lightest states can be estimated as  
$m_{ee}^{(2)} = (5 \cdot 10^{-7} - 10^{-5})$ eV  and 
$m_{ee}^{(1)}  < 2 \cdot 10^{-3}$ eV.   

\begin{figure}[!t]
\epsfxsize=60mm
\centerline{\epsfbox{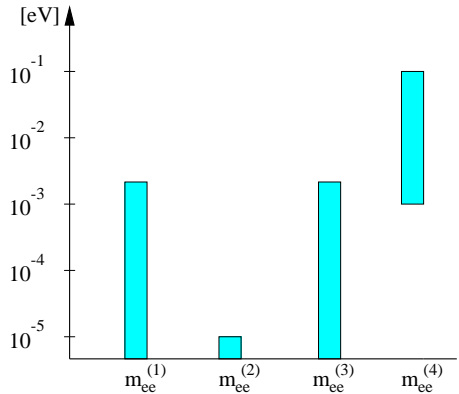}}
\caption{Contributions to $m_{ee}$ from different mass eigenstates 
in the $4\nu$ scheme with small flavor mixing and mass hierarchy. 
\label{cstates10}}
\end{figure}

Thus,  the \bbmass (see fig. \ref{cstates10}) can be  dominated by the
contribution 
of the heaviest state which can reach  
$m_{ee} \approx m_{ee}^{(4)} \sim 5 \cdot 10^{-2}$ eV.

The contribution  depends strongly 
on the mixing angle $\sin^2 2 \theta_{e \tau}$. 
Short baseline experiments
(such as the rejected
short baseline 
neutrino oscillation proposal  
TOSCA) could in principle test the region of 
large masses $m_{HDM} \simeq 10$ eV down to 
$\sin^2 2 \theta_{e \tau}=10^{-3}$, which correspond  to an improvement of 
the upper bound on $m_{ee}$ by 1 - 2 orders of magnitude. 
Due to possible cancellations between the contributions no 
lower bound on $m_{ee}$ can be obtained from the present data.

Notice that the MINIBOONE experiment will probe the mixing angle $\sin^2 2
\theta_{e \mu}$
down to $4 \cdot 10^{-4}$ eV and thus will 
check the LSND result. A confirmation of the LSND result 
will exclude this scheme.

\subsection{Scenario with two heavy degenerate  neutrinos}

\begin{figure}[!t]
\hbox to \hsize{\hfil\epsfxsize=8cm\epsfbox{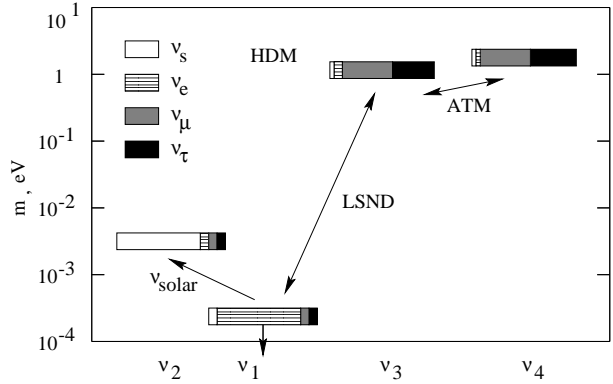}\hfil}
\caption{~~The pattern of the neutrino mass and mixing in the
scheme with two degenerate neutrinos and one sterile
component.
}
\label{fdeg}
\end{figure}

The main motivation for this scenario (see fig. \ref{fdeg}) 
is to explain the LSND result 
along with oscillation solutions of the solar and atmospheric neutrino
problems \cite{val2deg,moh2deg}. The masses are determined as 
\be
m_3 \approx m_4 \approx  \sqrt{\Delta m_{LSND}^2}, ~~~~   
m_2  \approx \sqrt{\Delta m_{\odot}^2}, ~~~~ m_1 \ll m_2~.   
\label{mass4}
\ee
The
neutrinos $\nu_{\mu}$ and $\nu_{\tau}$ are strongly 
mixed in the two heavy mass eigenstates $\nu_3$ and $\nu_4$,    
so that 
$\nu_{\mu} \leftrightarrow \nu_{\tau}$ oscillations 
solve the atmospheric neutrino problem.   
The two other neutrinos, $\nu_e$ and $\nu_s$, are weakly mixed in the two
lightest mass states and the resonance $\nu_e \rightarrow \nu_s$
conversion  solves the solar neutrino problem.  

The
two heavy neutrinos with masses  $m_{3} \approx m_{4}$  can be relevant
for cosmology,  their contribution to  a  hot dark matter component 
equals:  
\be 
m_{HDM}  = 2 m_{3} =  2\sqrt{\Delta m_{LSND}^2}~.  
\ee

In this scheme 
the
new element is the existence of two heavy degenerate states. 
Let us consider in details their contribution to $m_{ee}$ 
(the effect of the two lightest states is small).  
Using relations (\ref{mass4}) we can write this contribution as
\be 
m_{ee}^{(3)} + m_{ee}^{(4)} \approx  
(|U_{e3}|^2 +  |U_{e4}|^2 e^{i\phi_{34}}) \sqrt{\Delta m_{LSND}^2} 
\label{twostates1}
\ee
and $\phi_{34} \equiv \phi_{4} - \phi_{3}$ is the relative phase of the
$\nu_3$ and $\nu_4$  masses.  
Let us express the masses in eq. (\ref{twostates1}) in terms of 
oscillation parameters. In short base-line experiments 
the only oscillation phases which enter are the ones between 
heavy states and light states. 
One can neglect the oscillation phase  between the two light states  
which is determined by $\Delta m^2_{\odot}$ and the phase 
between the two heavy states which is determined 
by $\Delta m^2_{atm}$. In this case the 
oscillations are reduced to two neutrino 
oscillations with a phase determined by $\Delta m^2_{LSND}$  and the width 
for the $\nu_e \leftrightarrow \nu_{\mu}$ channel: 
\be
\sin^2 2\theta_{e \mu} = 
4 |U_{e3}^* U_{\mu 3} + U_{e 4}^* U_{\mu 4}|^2~. 
\label{emu}
\ee
Let us consider two extreme situations: suppose an  
admixture of the $\nu_e$ flavor 
in one of the heavy states is much larger than in the other one, 
{\it e.g.} $|U_{e3}| \gg |U_{e 4}|$, then 
$\sin^2 2\theta_{e \mu} = |U_{e3}|^2 |U_{\mu 3}|^2$,  and 
therefore $|U_{e3}|^2 = \sin^2 2\theta_{e \mu}/|U_{\mu 3}|^2$.  
In this case   we get from eq. (\ref{twostates1})
$m_{ee}^{(3)} + m_{ee}^{(4)} \approx |U_{e3}|^2 m_3$, and consequently,  
\be
m_{ee}^{(3)} + m_{ee}^{(4)} \approx
\frac{\sin^2 2\theta_{e \mu}}{|U_{\mu 3}|^2} \sqrt{\Delta m_{LSND}^2}~ .
\label{twostates}
\ee
Since
$|U_{\mu 3}|^2 \sim 0.5$ is determined by atmospheric 
neutrino oscillations, taking $\sin^2 2\theta_{e \mu} 
= 2 \cdot 10^{-3}$ and $\sqrt{\Delta m_{LSND}^2} = 1$ eV we find  
$m_{ee}^{(3)} + m_{ee}^{(4)} \approx 10^{-3}$ eV.

Let us now take 
$U_{e3}^* U_{\mu 3} \approx U_{e 4}^* U_{\mu 4}$, then 
$\sin^2 2\theta_{e \mu} = 16 |U_{e3}^* U_{\mu 3}|$ 
and $ m_{ee}^{(3)} + m_{ee}^{(4)} \approx
\sin^2 2\theta_{e \mu}  \sqrt{\Delta m_{LSND}^2}/2$,
provided that the  two contributions are in phase. 
This result is two times smaller than the result in the previous case.

For the $\nu_e \leftrightarrow \nu_{e}$ channel we find 
the depth of oscillations 
\be
\sin^2 2\theta_{e e} = 4 U_+ (1 - U_+) \approx 4 U_+, 
\label{eee}
\ee
where $U_+ \equiv |U_{e3}|^2 +  |U_{e 4}|^2$. 
At the same time 
$m_{ee} \leq U_+ \sqrt{\Delta m_{LSND}^2}/2$, so that 
\be  
m_{ee}^{(3)} + m_{ee}^{(4)} \approx
\frac{1}{4} \sin^2 2\theta_{e e} \sqrt{\Delta m_{LSND}^2}. 
\ee
Using the BUGEY bound on $\sin^2 2\theta_{e e}$
we get $m_{ee}^{(3)} + m_{ee}^{(4)} \lsim 10^{-2}$ eV. 

Since cancellations may show up,  no lower bound can be  obtained.

The same combination  of neutrino mixing
matrix elements (\ref{eee}) determines 
the $\nu_e$ mode of oscillations in atmospheric neutrinos. 
It will lead to an overall suppression of the number of the $e$-like events.

The contributions from the light states are similar  to those in the
$3\nu$ case (see sect. 3.1): 
$m_{ee}^{(2)} = (5 \cdot 10^{-7} - 10^{-5})$ eV and   
$m_{ee}^{(1)} \ll 2 \cdot 10^{-3}$ eV.   

\begin{figure}[!t]
\epsfxsize=60mm
\centerline{\epsfbox{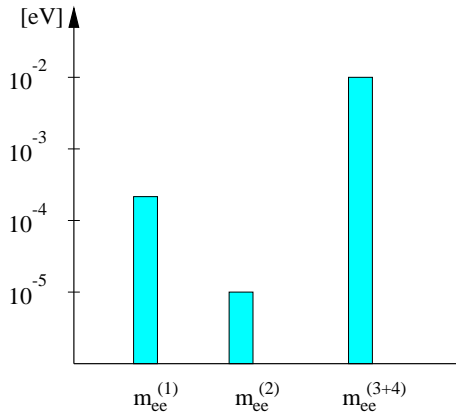}}
\caption{Contributions to  $m_{ee}$ 
from different mass eigenstates 
for the $4\nu$ scheme with two degenerate pairs of neutrinos and normal mass 
hierarchy.
\label{cstates11}}
\end{figure}

Summing up all the contributions we get that the
\bbmass can be at most $(few) \times 10^{-2}$ eV
being dominated by the 
contribution of the heavy states at the upper bound (fig.~\ref{cstates11}).  
A coincidence of a \znbb decay signal
in this range with a confirmation of the LSND oscillations by MINIBOONE
can  be considered as a hint for this scheme. 
At the same time, since cancellation 
between different  
contributions can show up, no lower bound on $m_{ee}$ exists.
Thus, a non-observation of \znbb decay of the order of magnitude 
$(few) \cdot 10^{-2}$ eV does not rule out the scheme.\\

\begin{figure}[!t]
\hbox to \hsize{\hfil\epsfxsize=8cm\epsfbox{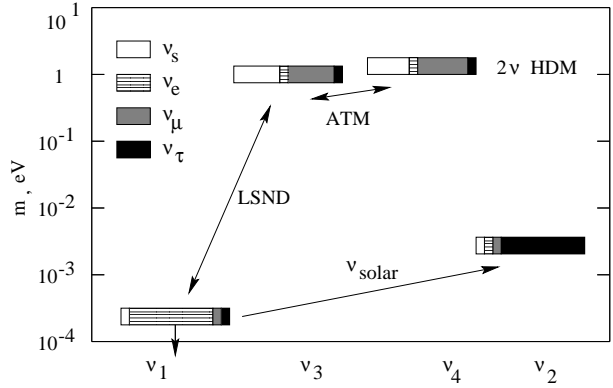}\hfil}
\caption{~~The neutrino masses and mixing
in the ``Grand Unification" scenario.
}
\label{fgu}
\end{figure}

A similar situation appears in the 
``Grand Unification"  scenario \cite{ptv,jossmi}
 which is characterized by 
strong mixing of $\nu_{\mu}$ and $\nu_s$ in the two heavy states and 
mixing of $\nu_{e}$ and $\nu_{\tau}$ in the two light states (fig. \ref{fgu}). 
Here the atmospheric neutrino problem is 
solved by $\nu_{\mu}
\leftrightarrow \nu_s$ oscillations whereas the solar neutrino data  
are explained by   $\nu_{e} - \nu_{\tau}$ conversion. 

\subsection{Scenario with inverse mass hierarchy}

The mass hierarchy in the two schemes with two pairs of  
states with small splitting can be inverse. 
In the first case, $\nu_e$ and $\nu_s$ flavors are concentrated in the 
two heavy states $\nu_3$ and  $\nu_4$, whereas $\nu_{\mu}$ 
and $\nu_{\tau}$ are in the two light states. The dominating contribution 
comes from the third state which almost coincides with 
$\nu_e$: 
\be
m_{ee} \approx m_{ee}^{(3)} \approx \sqrt{\Delta m_{LSND}^2}
\approx 0.4 - 1 {\rm eV}.  
\ee
Thus in the context of this scheme the double beta decay searches 
check immediately the LSND result, and in fact,  already existing 
data disfavor the scheme.

Another possibility of the inverse hierarchy is that the 
$\nu_e$ and $\nu_{\tau}$ flavors are concentrated in the heavy states, 
whereas $\nu_{\mu}$ and $\nu_s$ are in the pair of light 
mass states whose splitting  leads to 
the atmospheric neutrino oscillations. The situation is similar to that 
for the $3\nu$ scheme with inverse hierarchy (see sect. 7) with 
the only difference that $\Delta m_{atm}^2$ 
should be substituted by  $\Delta m_{LSND}^2$: 
\be
m_{ee}^{(3)}+ m_{ee}^{(4)} \simeq \sqrt{\Delta m^2_{LSND}}(\sin^2
\theta_{\odot}
+ e^{i \phi_{23}} \cos^2 \theta_{\odot})~.
\label{threefour}
\ee
The third and the fourth mass eigenstates give   
the dominating contributions.  
Thus the expected interval for the total effective mass 
is 
\be
m_{ee} \approx \sqrt{\Delta m^2_{LSND}}(\cos 2\theta_{\odot} - 1). 
\label{threefour1}
\ee
This interval can be probed already by existing experiments,  although
for large mixing angle solutions of the solar neutrino problem 
(LMA, LOW, VO) strong cancellation can occur. 

\section{Discussion and Conclusions}

We have performed a general analysis of the dependence of the effective
Majorana mass on the oscillation and non-oscillation parameters. 
Systematic studies of contributions from the individual mass eigenstates 
have been performed. We also have considered future developments
in view of forthcoming oscillation results. 
A systematic study of predictions from various schemes allows us to  
compare  these predictions and to conclude on implications 
of future double beta decay searches.

\begin{figure}[!t]
\epsfxsize=150mm
\centerline{\epsfbox{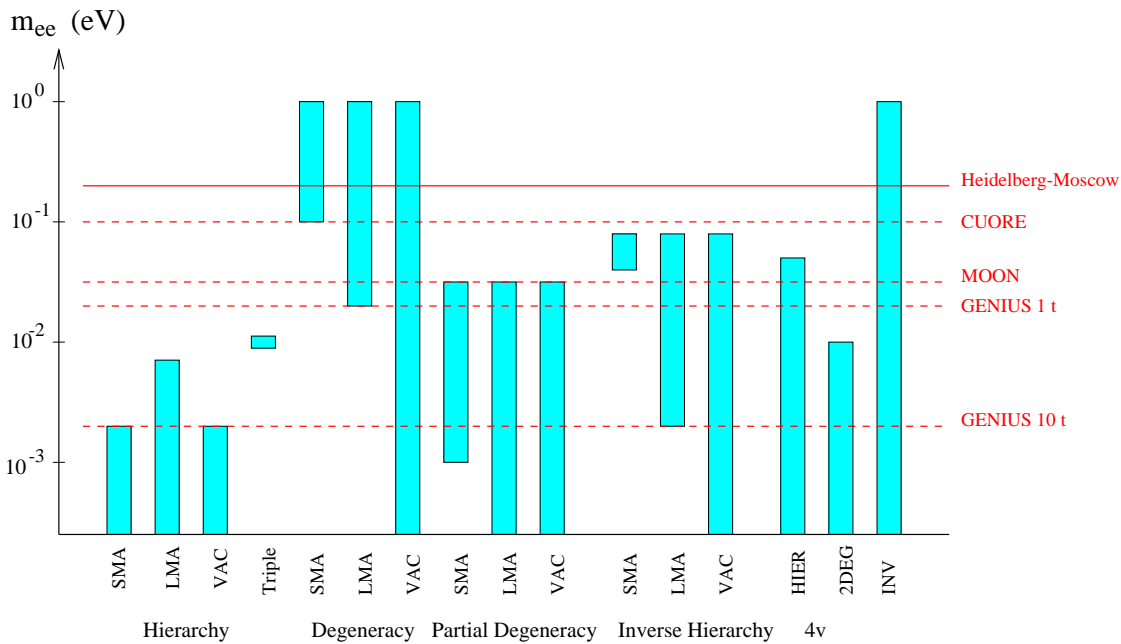}}
\caption{
Summary of expected values for $m_{ee}$ in the different schemes discussed
in this paper.
The expectations are compared with the recent neutrino mass limits 
obtained from the Heidelberg--Moscoscow \protect{\cite{hdmo}}
experiment as well as the expected 
sensitivities for the CUORE \protect{\cite{cuore}}, MOON 
\protect{\cite{moon}}
proposals 
and the 1 ton and 10 ton proposal of GENIUS \protect{\cite{gen}}.
\label{cstatessum}}
\end{figure}

In fig.~\ref{cstatessum} we summarize the predictions for $m_{ee}$ in various 
schemes considered in this paper.
We also show the present upper bound of \znbb decay experiments
\cite{hdmo} and regions of sensitivity 
which can be reached in future double beta decay experiments. 
Future double beta decay projects such as GENIUS \cite{Kla98,gen,gen2}, 
CUORE \cite{cuore}, MOON \cite{moon} will lead to
a significant improvement of the sensitivity. The most ambitious and at the 
same time most realistic
project, 
GENIUS, 
will test $m_{ee}$ down to $2 \cdot 10^{-2}$ eV in the one ton version with 
one year of measurement time and down to $2 \cdot 10^{-3}$ eV in the 
10 ton version 
with 10 years of measurement time.


According to figure \ref{cstatessum} there are two key scales 
of $m_{ee}$, which will allow one to discriminate  among 
various schemes: $m_{ee} \sim 0.1$ eV and $m_{ee} = 0.005$ eV. 

1).  If future searches 
will show that $m_{ee} >  0.1$ eV,  then the  schemes which 
will survive are those with neutrino mass degeneracy or 
$4 \nu$ schemes with inverse mass hierarchy. All other schemes 
will be excluded.

2). For masses in the interval $m_{ee} = 0.005 - 0.1$ eV,  
possible schemes include: $3\nu$ schemes with partial degeneracy, 
triple maximal scheme, $3\nu$ schemes with inverse mass hierarchy
and $4 \nu$ scheme with one heavy ({\cal O} (1 eV))  neutrino. 

3). If $m_{ee} < 0.005$ eV, the schemes which survive are 
$3\nu$ schemes with mass hierarchy, schemes with partial 
degeneracy, and the $4 \nu$ schemes with normal hierarchy. 
The schemes with degenerate spectrum and inverse mass hierarchy will be
excluded, unless large mixing allows for strong cancellations. 
For $m_{ee} < 0.001$ eV this applies also for schemes with a partial 
degenerate spectrum, again unless large mixing occurs.  

Future oscillation experiments will significantly reduce the uncertainty 
in predictions for $m_{ee}$  
and therefore modify implications of 
\znbb decay searches. Before a new generation of \znbb decay
experiments will start to
operate we can  expect that 

\begin{itemize}

\item
The solution of the solar neutrino problem will be identified. 
Moreover, $\Delta m^2_{\odot}$ and $\sin^2 2\theta_{\odot}$ 
will be determined with better accuracy. In particular, 
in the case of the solutions with large mixing (LMA, LOW, VO)  
the deviation of $1 - \sin^2 2\theta_{\odot}$ from zero can be established.

\item
The dominant channel of the atmospheric neutrino oscillation 
($\nu_{\mu} - \nu_{\tau}$ or $\nu_{\mu} - \nu_{s}$)  
will be identified. The mass $\Delta m^2_{atm}$ will be measured with 
better precision. 
 
\item 
A stronger bound on the  element $U_{e3}$ will be obtained or  it will be
measured  
if oscillations of electron neutrinos to tau neutrinos
driven by $\Delta m^2_{atm}$ 
will be discovered in the atmospheric neutrino or LBL experiments. 

\item
The 
LSND result will be checked by MINIBOONE.  

\end{itemize}

In the following we summarize possible consequences of these oscillation 
results. The conclusions obtained are not always stringent enough to 
exclude or prove any of the solutions in the whole parameter space.
In these cases we use phrases like ``favor'' or disfavor'' to describe 
the situation.

1). Let us first comment on how the identification of the solution 
of the solar neutrino problem will modify implications 
of the \znbb decay searches.


\begin{itemize}

\item
If the SMA solution of the solar neutrino problem turns out to be 
realized in nature, a value of $m_{ee}>0.2$ eV 
will imply a completely degenerate neutrino 
mass spectrum or schemes with inverse mass hierarchy. The measured value
of $m_{ee}$ will coincide with $m_1$ and will fix 
the absolute mass scale in the neutrino sector. A confirmation of this 
conclusion can  be obtained from the CMB experiments MAP and Planck, if
the degenerate neutrino mass is larger than $\simeq 0.1$ eV.

For lower values, $m_{ee}= 2 \cdot 10^{-3} - 10^{-2}$ eV, a   
scheme with partially degenerate spectrum will be favored. 
Again, we have $m_{ee}=m_1$ and the mass scale can be 
fixed.

For even lower mass values:  $m_{ee}< 2 \cdot 10^{-3}$ eV,  
or, after MINOS improved the bound on $m_{ee}^{(3)}$, 
$m_{ee}\lsim 4 \cdot 10^{-4}$ eV,
with the contribution $m_{ee}^{(3)}$ a new parameters enters, which for larger 
$m_{ee}$ could be neglected.
Thus it will be impossible to
quantify the contribution of each single state to $m_{ee}$, unless
$m_{ee}^{(3)}$ will be fixed in atmospheric or LBL oscillations.

\item
If the LMA solution of the solar neutrino deficit turns out to be 
realized in nature, a value $m_{ee}> 2 \cdot 10^{-2}$ eV will 
testify for a scenario with degenerate mass spectrum.
A confirmation of this result 
 will be obtained from the CMB experiments MAP and Planck, if
the degenerate neutrino mass is larger than $\simeq 0.1$ eV.
Using the mixing angle determined in solar neutrino 
experiments the range for the absolute mass scale can be determined 
from $m_{ee}$ 
according to fig.~\ref{smix4}.   
A value of $m_{ee}< 2 \cdot 10^{-2}$ eV will favor schemes with  
partial 
degeneracy or hierarchical spectrum. As soon as $m_{ee} <2 \cdot 10^{-3}$
eV $m_{ee}^{(3)}$ becomes important and enters as a new parameter and 
it will be difficult to reconstruct the type 
of hierarchy. 

\item
If the LOW or VO solution is the solution of 
the solar neutrino 
problem, the situation is similar to the MSW LMA case. The only difference is, 
that an observed \bbmass
$m_{ee}>2 \cdot 10^{-3}$ eV will imply a  partially or completely
degenerate
scheme.
Below this value the type of hierarchy can not be identified until bounds on 
$m_{ee}^{(3)}$ will be improved. 

\end{itemize}

For  schemes with inverse mass hierarchy  the situation can be more 
definite: 

\begin{itemize}

\item 
If the MSW SMA solution turns out to be true, a value of 
$m_{ee} =  \sqrt{\Delta m^2_{atm}} = 
(5 - 8) \cdot 10^{-2}$ eV is expected. 
This value coincides with $m_1 \simeq m_2$ and therefore will 
give the absolute mass scale.
For larger masses: $m_{ee} > 8 \cdot 10^{-2}$ eV the 
transition to a completely degenerate spectrum occurs.
  
\item 
If the MSW LMA, MSW LOW or vacuum  oscillation 
solution is realized, a value of
$m_{ee} = (0.02 - 8) \cdot 10^{-2}$ eV will testify for  
inverse mass hierarchy. 
The interval of expected values of $m_{ee}$ can be narrower once 
the deviation of $1 - \sin^2 2\theta$ from zero
will be measured in solar neutrino
experiments. 

For larger values of masses: $m_{ee}>8 \cdot 10^{-2}$ eV the scheme
approaches the 
degenerate case.

\end{itemize}

2). The discovery of a sterile neutrino will have significant impact on the
implications of the double beta decay searches.

The existence of a sterile neutrino can  be established by a confirmation
of the LSND result 
in MINIBOONE,  or by a proof of the $\nu_{e}\rightarrow \nu_s$ oscillations
solution of the solar neutrino problem by SNO, or by studies 
of the atmospheric neutrinos.   

For 4 $\nu$ scenarios the interpretation of 
the \znbb decay results is rather ambiguous.  
A value of $m_{ee}> (few)  \times 10^{-2}$ eV will 
favor the intermediate mass scale scenario,
while a value of $m_{ee}<10^{-3}$ eV will favor 
a scenario with two degenerate
pairs of neutrinos and normal mass hierarchy. A value of $m_{ee}>10^{-1}$ eV
will clearly disfavor a strongly hierarchical scheme with normal mass 
hierarchy and favor the cases of inverse hierarchy or degeneracy.
In all cases it will be difficult to disentangle the 
single contributions and to identify a specific spectrum.
Important input in this case may come from the CMB experiments MAP and Planck
by fixing the mass of the heaviest state.\\

3). $U_{e3}$: further searches for $\nu_e$ oscillations in 
atmospheric neutrinos,  LBL and reactor experiments will 
allow one to measure or further restrict this mixing element. 
This, in turn, will be important for sharpening the predictions 
for $m_{ee}$ especially in the schemes with strong mass hierarchy. 
\\

4). Matter effects and hierarchy: Studies of matter effects on neutrino
oscillations will allow to establish the type of mass hierarchy, which in
turn is of great importance for predictions of $m_{ee}$. \\

We can conclude from this summary, that in 3$\nu$ scenarios
any measurement of 
$m_{ee}> 2 \cdot 10^{-3}$ eV in \znbb decay 
(corresponding to the final sensitivity 
of the 10 ton version of GENIUS) will provide informations about
 the character of hierarchy
of the neutrino mass spectrum and in some cases also to
fix the absolute mass scale of  
neutrinos. For values of $m_{ee}<2 \cdot 10^{-3}$ eV no reconstruction of
the spectrum is possible until the contribution $m_{ee}^{(3)}$
will be fixed or bounded more stringent in atmospheric or LBL neutrino 
oscillations.
For four-neutrino scenarios it will be not that easy to fix the mass scale
of the neutrino sector. Crucial informations can be obtained from tests of the
LSND signal and cosmology.

As has been mentioned before, a non-zero \znbb decay rate always
implies a non-vanishing neutrino Majorana mass \cite{schech}.
Let us comment finally on possible ambiguities in the interpretation
of a positive signal in neutrinoless
double beta decay in terms 
of $m_{ee}$, in view of the existence of 
different alternative mechanisms, which could 
induce neutrinoless double beta decay, such as R-parity violating SUSY, 
right-handed currents, or leptoquarks. While no absolute unique method 
to identify the mechanism being responsible for neutrinoless double beta decay
exists, the following remarks can be done:

1). Many of the possible alternative contributions require new particles,
{\it e.g.} SUSY partners, leptoquarks, right-handed W bosons or neutrinos
having masses in or below the TeV range, which to date not have been 
observed. Thus one expects to observe  
effects of new particles at future high energy colliders as the LHC or
the NLC, giving independent informations on possible contributions to 
\znbb decay 
(keeping in mind an uncertainty in
nuclear matrix elements of about a factor of ${\cal O}(2)$).
Notice that the same new interactions mentioned here may induce effects in
neutrino ocillations and imply ambiguities in the interpretation of the data
also there (see e.g. \cite{bergm,valnp}).

2). Using  different source isotopes in 
different experiments and figuring out the values of 
\znbb decay 
nuclear matrix elements for different contributions may help to identify 
the dominant one. Also a future experiment being sensitive to angular 
correlations of outgoing electrons could be useful in the discrimination of 
different contributions.
Observing a positive signal in \znbb decay should encourage new experimental 
efforts to confirm the results. 

3).  Last but not least and as discussed in this paper, 
a non-zero \znbb decay signal 
can be related to  some experimental results
(both positive and negative)
in neutrino oscillations 
and cosmology. 
A coincident and non-contradictory identification of a single neutrino mass 
scheme from the
complementary results in such different experiments thus should be
respected
as a strong hint for this scheme.

In conclusion, after Super-Kamiokande has established large mixing in 
atmospheric 
neutrinos, the simplest neutrino spectrum with strong mass hierarchy and
small flavor mixing (which typically predicts an 
undetectable $m_{ee}$) is excluded. Now the neutrino mass spectrum can 
exhibit
any surprise: it can have a normal or inverse mass hierarchy, be partially
or completely degenerate. More than three mass eigenstates can be involved
in the mixing. In view of this more complicated situation a  detection of
a positive signal in future double beta decay searches seems to be
rather  plausible. We have shown  that for a given  oscillation pattern
any value of $m_{ee}$ is possible, which is still not excluded by the 
experimental bounds on the neutrinoless double beta decay half life limit.
(A lower bound on $m_{ee}$ appears in the case of inverse mass
hierarchy.) This means that even after all oscillations parameters will be
measured  no unique prediction of $m_{ee}$ can be derived. On the other hand
this means that double beta decay searches provides informations being
independent on informations obtained from oscillation experiments.
Combining the results of double beta decay and oscillation searches
offers a unique possibility to shed some light on the
absolute scale of the neutrino mass,
the type of hierarchy and the level of degeneracy of the spectrum. If we want
eventually to reconstruct the neutrino mass and flavor spectrum, further
searches for neutrinoless double beta decay with increased sensitivity
seem to be unavoidable.

{\bf Note added:} 
When this paper has been prepared for submission
the papers of ref.  \cite{spaet} appeared which discuss similar topics.

\end{document}